\documentclass[a4paper,11pt]{article}
\usepackage{amsmath}
\usepackage{amssymb}
\usepackage{graphicx}
\usepackage{parskip}
\usepackage{amsthm}
\usepackage{fancyhdr}
\usepackage{dsfont}
\usepackage{algorithm}
\usepackage{algorithmic}
\usepackage{hyperref}
\usepackage{bm}
\usepackage[authoryear]{natbib}

\theoremstyle{definition}

\newtheorem{rmk}{Remark}

\newcommand{\N}{ \mathbb{ N } }
\newcommand{\E}{ \mathbb{ E } }

\newcommand{\Prb}{ \mathbb{ P } }

\newcommand{\bd}{\begin{displaymath}}
\newcommand{\ed}{\end{displaymath}}
\newcommand{\be}{\begin{equation}}
\newcommand{\ee}{\end{equation}}

\title{Robust model selection between population growth and multiple merger coalescents}

\author{Jere Koskela \\
	\texttt{j.koskela@warwick.ac.uk}\\
	\small Department of Statistics\\
	\small University of Warwick \\
	\small Coventry, CV4 7AL \\
	\small United Kingdom
	\and
	Maite Wilke Berenguer \\
	\texttt{maite.wilkeberenguer@ruhr-uni-bochum.de} \\
	\small Fakult\"at f\"ur Mathematik  \\
	\small Ruhr Universit\"at Bochum\\
	\small Universit\"atstra{\ss}e 150, 44780 Bochum, \\
	\small Germany
}
\date{\today}

\begin{document}

\maketitle

\begin{abstract}
We study the effect of biological confounders on the model selection problem between Kingman coalescents with population growth, and $\Xi$-coalescents involving simultaneous multiple mergers.
We use a low dimensional, computationally tractable summary statistic, dubbed the \emph{singleton-tail statistic}, to carry out approximate likelihood ratio tests between these model classes.
The singleton-tail statistic has been shown to distinguish between them with high power in the simple setting of neutrally evolving, panmictic populations without recombination.
We extend this work by showing that cryptic recombination and selection do not diminish the power of the test, but that misspecifying population structure does.
Furthermore, we demonstrate that the singleton-tail statistic can also solve the more challenging model selection problem between multiple mergers due to selective sweeps, and multiple mergers due to high fecundity with moderate power of up to 30\%.
\end{abstract}

\section{Introduction} \label{intro}

The Kingman coalescent \citep{K82,K82b,Ki82c, H1983a,H1983b, T1983} models ancestral relations of samples from large populations as random, binary trees, and is an important tool for predicting genetic diversity.
A central assumption of the Kingman coalescent is low variance of family sizes, so that large populations always consist of many relatively small families.
Violations of this assumption call for models with infinite variance family sizes, and lead to so called $\Lambda$-coalescents, which allow more than two lineages to merge to a common ancestor simultaneously \citep{DK1999, P1999, S1999}.

There is growing evidence that $\Lambda$-coalescents are an appropriate model for organisms with high fecundity coupled with a skewed offspring distribution \citep{B1994,A2004,EW2006,SW2008,HP2011, BBS2011, SBB2013, TL14}.
Consequently, development of statistical techniques for distinguishing the Kingman coalescent from $\Lambda$-coalescents has also been an active area of research; see \citep{EBBF15, K18}, and references therein.
In particular, attention has focused on distinguishing $\Lambda$-coalescents from Kingman coalescents with population growth, because both classes of models predict an excess of singletons (mutations only carried by one individual in a sample of DNA sequences) relative to the standard Kingman coalescent under the infinitely many sites model of mutation \citep{W75}.

\cite{K18} introduced a simple, two-dimensional summary statistic, referred to here as the \emph{singleton-tail statistic}, which distinguishes between these model classes with high power even from a data set consisting of 500 samples from bi-parental, diploid organisms sequenced at around 10 unlinked chromosomes.
The correct model could be selected with high power without knowing the population-rescaled mutation rate, provided it is was not very low (see also \cite[Supporting Information 12]{EBBF15}).
In this paper we investigate  the impact of other confounders on the prospect of discriminating between these models based on the singleton-tail statistic, again in the bi-parental, diploid setting.
In particular, we will focus on each of
\begin{enumerate}
\item weak natural selection modelled by the Ancestral Selection Graph \citep{KN97, NK97, DK99, BLW16},
\item crossover recombination within chromosomes modelled by the Ancestral Recombination Graph \citep{H1983a, GM97, DK99, BBE13a},
\item population structure modelled by the structured coalescent \citep{H97, VL06, Eldon09}.
\end{enumerate}
We will demonstrate that the presence or absence of the first two has minimal effect on the performance of the hypothesis test developed in \citep{K18}, while population structure is a significant counfounder that must be correctly incorporated into the model.

There are four parental copies of each chromosome involved in each merger in the diploid, bi-parental setting, allowing for up to four simultaneous mergers.
Hence the models considered in this paper are actually $\Xi$-coalescents \citep{S2000, MS2001} which allow simultaneous multiple mergers, despite the fact that the population will be assumed to reproduce in a fashion consistent with the more restrictive $\Lambda$-coalescent permitting only one multiple merger at a time.

We also use the singleton-tail statistic to distinguish two classes of $\Lambda$-coalescents: those arising from high fecundity reproduction, and those arising from selective sweeps \citep{DS05}.
This problem is more challenging than a null hypothesis consisting of Kingman coalescents with population growth, because the marginal coalescent process at each chromosome can be identical under the two hypotheses.
However, high fecundity reproduction results in positively correlated coalescence times between unlinked chromosomes, whereas unlinked chromosomes are independent under the selective sweep model.
The positive correlation results in increased sampling variance of the singleton-tail statistic, which yields tests with moderate statistical power of up to $30\%$.

The rest of the paper is organised as follows.
In Section \ref{single_tail} we recall the singleton-tail statistic of \citep{K18} as well as the associated hypothesis test for model selection.
Section \ref{alt_models} presents a unified, diploid coalescent model incorporating high fecundity reproduction and population growth, as well as the three confounders of weak selection, crossover recombination, and discrete spatial structure.
Models with only population growth or high fecundity reproduction, as well as any desired subset of confounders, can be recovered as special cases.
Section \ref{robustness} provides simulation studies on the effect of each of the three confounders on the sampling distribution of the singleton-tail statistic, as well as the associated hypothesis test.
In Section \ref{sweeps} we introduce a different model in which rapid selective sweeps result in multiple mergers acting locally on the genome, and investigate whether the singleton-tail statistic can distinguish it from the $\Xi$-coalescent introduced in Section \ref{alt_models}.
Section \ref{discussion} concludes with a discussion.

\section{The singleton-tail statistic}\label{single_tail}

Suppose a sample of $n \in \N$ DNA sequences from a single chromosome is available, and that derived mutations can be distinguished from ancestral states.
Let $[n]:=\{1, \dots, n\}$, and let $\xi_i^{(n)}$ be the number of sites at which a mutant allele appears $i \in [ n - 1 ]$ times. 
Then
\begin{equation*}
\bm{\xi}^{(n)} := \left(\xi_1^{(n)}, \ldots , \xi_{n-1}^{(n)} \right)
\end{equation*}
is  the \emph{unfolded site-frequency spectrum} (SFS).
If mutant and ancestral types cannot be distinguished, the \emph{folded} spectrum $\bm{ \eta }^{ ( n ) } := ( \eta_1^{ ( n ) }, \ldots, \eta_{ \lfloor n / 2 \rfloor }^{ ( n ) } )$ \citep{Fu1995} is used instead, where
\begin{equation*}
\eta_i^{(n)} :=  \frac{\xi_i^{(n)} + \xi_{n-i}^{(n)}}{1  +  \delta_{i,n-i}}, \quad 1 \le i \le \lfloor n/2 \rfloor,
\end{equation*}
and $\delta_{i,j} = 1$ if $i = j$, and is zero otherwise. 
Let $\bm{\zeta}^{(n)} := (\zeta_1^{(n)}, \ldots , \zeta_{n-1}^{(n)})$ be the normalised unfolded SFS, whose entries are given by $\zeta_i^{(n)} := \xi_i^{(n)}/|\bm{\xi}^{(n)}|$, where $|\bm{ \xi }^{(n)}| := \xi_1^{(n)} + \cdots + \xi_{n-1}^{(n)}$ is the total number of segregating sites, and with the convention that $\bm{\zeta}^{(n)} = \bm{0}$ if there are no segregating sites.

Now, for any $k \in [ n - 1 ]$ define the \emph{lumped tail} of the SFS as
\begin{equation*}
\overline{ \zeta }^{ ( n ) }_k := \sum_{ j = k }^{ n - 1 } \zeta_j^{ ( n ) },
\end{equation*}
and consider the summary statistic $( \zeta_1^{ ( n ) }, \overline{ \zeta }_k^{ ( n ) } )$ for some fixed $k$.
Data from multiple chromosomes is incorporated by averaging: if $L$ unlinked chromosomes are available, then the singleton-tail statistic is
\begin{equation*}
( \zeta_{1, L}^{ ( n ) }, \overline{ \zeta }_{k, L}^{ ( n ) } ) := \frac{ 1 }{ L }\sum_{ j = 1 }^L ( \zeta_1^{ ( n ) }( j ), \overline{ \zeta }_k^{ ( n ) }( j ) ),
\end{equation*}
where $( \zeta_1^{ ( n ) }( j ), \overline{ \zeta }_k^{ ( n ) }( j ) )$ denotes the singleton class and lumped tail computed from the $j^{\text{th}}$ chromosome.

For two classes of models $\Theta_0$ and $\Theta_1$, the likelihood ratio test statistic is
\begin{equation*}
\frac{ \sup_{ \Pi \in \Theta_1 } P^{ \Pi }( \zeta_{1, L}^{ ( n ) }, \overline{ \zeta }_{k, L}^{ ( n ) } ) }{ \sup_{ \Pi \in \Theta_0 } P^{ \Pi }( \zeta_{1, L}^{ ( n ) }, \overline{ \zeta }_{k, L}^{ ( n ) } ) },
\end{equation*}
where $P^{ \Pi }$ denotes the sampling distribution of the singleton-tail statistic under coalescent $\Pi$.
A corresponding hypothesis test of size $\omega \in ( 0, 1 )$ given an observed value of the singleton-tail statistic is
\begin{equation}\label{test}
\Phi( \zeta_{1, L}^{ ( n ) }, \overline{ \zeta }_{k, L}^{ ( n ) } ) = \begin{cases}
0 &\text{if } \frac{ \sup_{ \Pi \in \Theta_1 } P^{ \Pi }( \zeta_{1, L}^{ ( n ) }, \overline{ \zeta }_{k, L}^{ ( n ) } ) }{ \sup_{ \Pi \in \Theta_0 }  P^{ \Pi }( \zeta_{1, L}^{ ( n ) }, \overline{ \zeta }_{k, L}^{ ( n ) } ) } \leq q_{ \omega } \\
1 &\text{if } \frac{ \sup_{ \Pi \in \Theta_1 } P^{ \Pi }( \zeta_{1, L}^{ ( n ) }, \overline{ \zeta }_{k, L}^{ ( n ) } ) }{ \sup_{ \Pi \in \Theta_0 } P^{ \Pi }( \zeta_{1, L}^{ ( n ) }, \overline{ \zeta }_{k, L}^{ ( n ) } ) } > q_{ \omega }
\end{cases},
\end{equation}
where $\Phi( \zeta_{1, L}^{ ( n ) }, \overline{ \zeta }_{k, L}^{ ( n ) } ) = 1$ corresponds to rejecting the null hypothesis $\Theta_0$, and $q_{ \omega }$ is the quantile
\begin{equation*}
q_{ \omega } := \inf\left\{ q \geq 0 : \sup_{ \Pi \in \Theta_0 } \Prb^{ \Pi }\left( \frac{ \sup_{ \Pi \in \Theta_1 } P^{ \Pi }( \zeta_{1, L}^{ ( n ) }, \overline{ \zeta }_{k, L}^{ ( n ) } ) }{ \sup_{ \Pi \in \Theta_0 } P^{ \Pi }( \zeta_{1, L}^{ ( n ) }, \overline{ \zeta }_{k, L}^{ ( n ) } ) }  \geq q \right) \leq \omega \right\}.
\end{equation*}
The sampling distribution $P^{ \Pi }$ and the quantile $q_{ \omega }$ are both intractable, but can easily be approximated by simulation due to the low dimensionality of the singleton-tail statistic to obtain an implementable hypothesis test with approximate size $\omega$ \cite{K18}.
In particular, we consider the hypotheses
\begin{align}
\Theta_0 := \{& \text{Kingman coalescent with exponential growth at population-rescaled rate} \nonumber \\
&\gamma \in \{ 0, 0.1, 0.2, \ldots, 0.9, 1, 1.25, 1.5, 2, 2.5, 3, 3.5, 4, 5, 6, \ldots, 19, 20, \nonumber \\
&\phantom{\gamma \in \{}25, 30, 35, 40, 50, 60, \ldots, 990, 1000 \} \}, \nonumber \\
\Theta_1 := \{&\text{Beta}(2 - \alpha, \alpha)\text{-}\Xi\text{-coalescents with } \alpha \in \{ 1, 1.025, \ldots, 1.975, 2\} \}, \label{theta_1}
\end{align}
In brief, data is simulated under both $\Theta_0$ and $\Theta_1$, and kernel density estimates (KDEs) $\hat{ P }^{ \Pi }$ of the intractable sampling distributions $P^{ \Pi }$ are obtained for $\Pi \in \Theta_0$ and $\Pi \in \Theta_1$.
These KDEs, along with more simulated data, can be used to accurately approximate the intractable quantile $q_{ \omega }$, yielding an implementable hypothesis test.
Our KDEs were obtained using the \texttt{kde} function in the \texttt{ks} package (version 1.10.4) in \texttt{R} under default settings.
In particular, this method uses truncated Gaussian kernels, and determines bandwidths using the SAMSE estimator \cite[equation (6)]{Duong03}.
\vskip 11pt
\begin{rmk}
The null hypothesis in \cite{K18} was broader and included algebraic population growth, in addition to exponential.
However, results in \cite{K18} showed that the two growth models resulted in very similar sampling distributions for the singleton-tail statistic, and hence we focus on the exponential growth model.
\end{rmk}

Simulating data in order to approximate the test \eqref{test} requires specification of the cutoff $k$ for the lumped tail of the singleton-tail statistic, as well as of the mutation rate $\theta$.
Sensitivity analyses conducted in \cite{K18} showed that the test was highly insensitive to the choice of $k$ provided $k \gtrapprox 6$, as well as to misspecification of the mutation rate by up to a factor of ten.
We fix $k = 15$ throughout, and use the known, true mutation rate in our simulation studies.
For biological data sets with an unknown mutation rate, the analysis in \cite{K18} demonstrated that it is sufficient to use the generalised Watterson estimator, 
\begin{equation*}
\hat{ \theta } = | \bm{ \xi }^{ ( n ) } | / \E^{ \Pi }[ T^{ ( n ) } ],
\end{equation*}
where $\E^{ \Pi }[ T^{ ( n ) } ]$ is the expected branch length from $n$ leaves under coalescent $\Pi$.

\section{An umbrella model}\label{alt_models}

In this section we describe a general class of models incorporating diploidy, bi-parental, high fecundity reproduction, population growth, weak natural selection, population structure in a discrete geography, and crossover recombination.
This generalises both the Ancestral Influence Graph \cite{DK99}, as well as time-inhomogeneous multiple merger coalescents \cite{M2002, MHAJ18}.
Models with any subset of the above forces can be recovered as special cases.

Consider a geography of $D$ demes, with the population size on deme $i \in [ D ]$ at time $t$ given by $2 M_N^{ ( i ) }( t )$, where $N$ is a scaling parameter.
We also define the total population size
\begin{equation*}
2 M_N( t ) := \sum_{ i = 1 }^D 2 M_N^{ ( i ) }( t ),
\end{equation*}
and the shorthand $M_N := M_N( 0 )$.

Each individual carries a diploid genome consisting of $L \in \N$ pairs of unlinked chromosomes.
Each chromosome carries one of $K \in \N$ alleles, identified with $[ K ]$, which are acted upon by natural selection. 
In addition, each chromosome is identified with the unit interval $[ 0, 1 ]$, on which neutral mutations and crossover recombination take place.
For definiteness, we assume that the selective allele is fully linked to the left end of the neutral interval.

The populations evolve in discrete time with non-overlapping generations. 
At each time $t$, the individuals in each deme form pairs uniformly at random.
The pairs are ordered in a fixed but arbitrary way, and pair $j$ on deme $i$ has a random number of offspring denoted by $\nu_j^{ ( i ) }( t ) + \beta_j^{ ( i ) }( t )$.
The two summands will be associated with neutral reproduction and natural selection, respectively.
As such, the distribution of $\beta_j^{ ( i ) }( t )$ will depend on the alleles of the two parents, though this dependence is suppressed for legibility.
Likewise, we will frequently suppress the time-dependence in the family sizes, and write $\nu_j^{ ( i ) }$ and $\beta_j^{ ( i ) }$.
For future convenience, we define $\tilde{ \beta }_j^{ ( i ) }( t )$ as the random number of selective offspring that pair $j$ on deme $i$ at time $t$ would have had if they carried the fittest possible combination of alleles.

The neutral offspring vectors $( \nu_1^{ ( i ) }, \ldots, \nu_{ M_N^{ ( i ) }( t ) }^{ ( i ) } )$ are assumed to be exchangeable, and independent across demes as well as time steps.
The selective offspring vectors  $( \beta_1^{ ( i ) }, \ldots, \beta_{ M_N^{ ( i ) }( t ) }^{ ( i ) } )$ are independent across demes and time steps.
In addition, both vectors on each deme are assumed to satisfy the almost sure constraint
\begin{equation*}
\sum_{ j = 1 }^{ M_N^{ ( i ) }( t ) } \nu_j^{ ( i ) } + \beta_j^{ ( i ) } \equiv 2 M_N^{ ( i ) }( t + 1 ).
\end{equation*}
Each offspring inherits one copy of each of its $L$ chromosome pairs from each of its parents.
Each inherited chromosome is a mosaic of the two chromosomes carried by the parent, with the number of recombination breakpoints having the Poisson distribution with parameter $r_N$, and each break point being uniformly distributed along the chromosome.
All of the Poisson and uniform random variables are independent of each other, as well as of the wider reproduction mechanism.
Each locus inherits its allele from the parental chromosome assigned to its leftmost  segment, with selective mutations happening independently at random with probability $\mu_N$.
Mutant types are drawn from a stochastic matrix $M = ( M_{ i j } )_{ i, j = 1 }^K$, where $M_{ i j }$ is the probability of a mutant locus having allele $j$ given its parent had allele $i$.

After the reproduction step is complete, a deterministic fraction $m_{ i j }^{ ( N ) }$ of children chosen uniformly at random from deme $i$ migrate to deme $j$, for each pair of demes.
These migration fractions are assumed to satisfy
\begin{equation}\label{reversible_migration}
M_N^{ ( i ) }( t ) \sum_{ j = 1 }^D m_{ i j }^{ ( N ) } \equiv \sum_{ j = 1 }^D M_N^{ ( j ) }( t ) m_{ j i }^{ ( N ) },
\end{equation}
for each $t \geq 0$, so that the population sizes of demes remain unchanged by migration.
For notational brevity we set $m_{ i i }^{ ( N ) } \equiv 0$.

We now reverse the direction of time, so that time $t \in \N$ corresponds to $t$ generations in the past in the model specified above.
For $n \in \N$ and $k \in \N$ let $( n )_k := n ( n - 1 ) \ldots ( n - k + 1 )$ denote the falling factorial, and define
\begin{align}
c_N^{ ( i ) }( t ) &:= \frac{ M_N^{ ( i ) }( t ) }{ 4 ( 2 M_N^{ ( i ) }( t - 1 ) )_2 } \E[ ( \nu_1^{ ( i ) } )_2 ], \nonumber \\
c_N &:= \frac{ 1 }{ 4 ( 2 M_N )_2  } \sum_{ i = 1 }^D M_N^{ ( i ) }( 1 ) \E[ ( \nu_1^{ ( i ) } )_2 ], \label{c_N_defn}
\end{align}
as the probability that two chromosomes sampled uniformly at random from deme $i \in [ D ]$ (resp.~the whole population) at time $t \in \N$ (resp.~$t = 0$) were born to a common family in the previous generation, made the same choice from two available parents, and also the same choice of chromosome within that parent.
In other words, $c_N^{ ( i ) }( t )$ is the probability of two time $t$ chromosomes on island $i$ merging to a common ancestor in one generation, while $c_N$ is the same probability for two chromosomes sampled uniformly from the whole population at time $t = 0$.

We make the following assumptions for each $i, j \in [ D ]$, and each $t \in ( 0, \infty )$, as $N \rightarrow \infty$, where each $\Lambda_i$ is a probability measure on $[ 0, 1 ]$, and each $\lambda_i( t )$ is a positive function bounded away from 0, with $\lambda_1( 0 ) + \ldots + \lambda_D( 0 ) = 1$ and $\inf_{ t \geq 0, i \neq j }\{ \lambda_i( t ) / \lambda_j( t ) \} > 0$, and $\gamma \in [ 0, 1 )$ and $C > 0$ are constant independent of $N$, $i$, and $t$:
\begin{align}
c_N &\rightarrow 0, \label{a4} \\
\inf_{ i \in [ D ], t \geq 0 }\{ M_N^{ ( i ) }( t ) \} &\rightarrow \infty, \label{a5}\\
\E[ ( \nu_1^{ ( i ) }( t ) )_2 ] &\sim C M_N^{ ( i ) }( t )^{ \gamma }, \label{a6} \\
\frac{ M_N^{ ( i ) }( \lfloor t / c_N \rfloor ) }{ M_N } &\rightarrow \lambda_i( t ), \label{a7} \\
\frac{ ( M_N^{ ( i ) }( \lfloor t / c_N \rfloor ) )_2 \E[ ( \nu_1^{ ( i ) } + \tilde{ \beta }_1^{ ( i ) } )_2 ( \nu_2^{ ( i ) } + \tilde{ \beta }_2^{ ( i ) } )_2 ] }{ ( 2 M_N^{ ( i ) }( \lfloor t / c_N \rfloor - 1 ) )_4 c_N } &\rightarrow 0, \label{a8} \\
\frac{ M_N^{ ( i ) }( \lfloor t / c_N \rfloor - 1 ) }{ c_N^{ ( i ) }( \lfloor t / c_N \rfloor ) } \Prb( \nu_1^{ ( i ) } > 2 M_N^{ ( i ) }( \lfloor t / c_N \rfloor - 1 ) x ) &\rightarrow \int_x^1 \frac{ \Lambda_i( dy ) }{ y^2 }, \label{a9} \\
\mu_N / c_N &\rightarrow \theta \in [ 0, \infty ), \label{a10} \\
r_N / c_N &\rightarrow \rho \in [ 0, \infty ), \label{a11} \\
m_{ i j }^{ ( N ) } / c_N &\rightarrow m_{ i j } \in [ 0, \infty ), \label{a12} \\
\E[ \tilde{ \beta }_1^{ ( i ) } ] / c_N &\rightarrow \sigma_i \in [ 0, \infty ), \label{a13} \\
\frac{ 1 }{ c_N } \sup_{ k \geq 1 }\left\{ \E\left[ \tilde{ \beta }_1^{ ( i ) } \left( \nu_1^{ ( i ) } + \sum_{ j = 1 }^{ M_N^{ ( i ) }( \lfloor t / c_N \rfloor ) } \tilde{ \beta }_j^{ ( i ) } \right)^k \right] \right\} &\rightarrow 0 \label{a14}.
\end{align}
\begin{rmk}
It is well known that if $\Lambda_i = \delta_0$, the Dirac delta-measure at 0, in \eqref{a9}, then the assumption is equivalent to
\begin{equation*}
\frac{ \E[ ( \nu_1^{ ( i ) } )_3 ] }{ 4 ( 2 M_N^{ ( i ) }( \lfloor t / c_N \rfloor - 1 ) )_2 c_N^{ ( i ) }( \lfloor t / c_N \rfloor ) } \sim \frac{ \left( \sum_{ j = 1 }^D \lambda_j( 0 )^{ \gamma + 1 } \right) \E[ ( \nu_1^{ ( i ) } )_3 ] }{ 16 M_N^2 \lambda_i( t )^{ \gamma + 1 } c_N } \rightarrow 0
\end{equation*}
for each $t \in ( 0, \infty )$ and $i \in [ D ]$, where the second representation follows from \eqref{a7} and
\begin{equation}\label{c_N_equivalence}
c_N^{ ( i ) }( \lfloor t / c_N \rfloor ) \sim \frac{ \lambda_i( t )^{ \gamma - 1 } }{ \sum_{ j = 1 }^D \lambda_j( 0 )^{ \gamma + 1 } } c_N,
\end{equation}
itself a consequence of \eqref{a6} and \eqref{a7}.
See \cite[Section 5]{MS2003} for details.
This assumption disallows multiple mergers in the limiting ancestry, which will thus only consist of isolated binary mergers.
Any other choice of $\Lambda_i$ will yield an ancestry with up to four simultaneous multiple mergers at each chromosome, corresponding to the four possible parental chromosomes involved in the forwards-in-time reproduction event, and thus produce ancestries described by a $\Xi$-coalescent.
See \cite[Section 6]{MS2003} for details of $\Xi$-coalescents arising out of diploid reproduction in this way.
\end{rmk}
\vskip 11pt
\begin{rmk}
Before showing that \eqref{a4} -- \eqref{a14} lead to the desired ancestral process, some intuition behind the role of each assumption is in order.
\eqref{a4} yields a limit process evolving in continuous time.
Assumptions \eqref{a5} -- \eqref{a7} ensure that the population sizes and time scales on demes are comparable.
The conditions on the relative population sizes $\lambda_i( t )$ are sufficient  to ensure finite waiting times between merger and migration events, and could be relaxed in specific examples.
For exponential population growth, they hold as long as the growth rates on all demes coincide.
For models in the domain of attraction of Kingman's coalescent, \eqref{a6} will typically hold with $\gamma = 0$, while e.g.~the $\operatorname{Beta}( 2 - \alpha, \alpha )$-coalescents of \cite{S2003} have $\gamma = 2 - \alpha$ (c.f.~\eqref{c_N_defn} and \cite[Lemma 13]{S2003}).
The $\gamma \in [ 0, 1 )$ condition ensures that \eqref{a4} and \eqref{a6} can hold simultaneously.
Conditions \eqref{a8} and \eqref{a9} are well known to be necessary and sufficient for a $\Lambda$-coalescent limit, resulting in no more than four simultaneous multiple mergers in the diploid, biparental setting. 
\eqref{a10} -- \eqref{a13} ensure that mutation, recombination, migration, and selection all take place on the coalescent time scale, while \eqref{a14} disallows multiple selective branching events, as well as simultaneous selective and neutral merger events.
\end{rmk}

The aim is to show that the ancestry of a sample from the above particle system converges to a structured, time-inhomogeneous $\Xi$-Ancestral Influence Graph \citep{DK99} as $N \rightarrow \infty$, when time is measured in units of $c_N$.
To establish this fact, we identify the limiting rates of coalescence, mutation, recombination, migration and branching due to selection, and show that these are the only dynamics which affect the ancestry of the process.
Specifically, that $r \leq 4$ simultaneous mergers of sizes $b_1, \ldots, b_r$, with $2 \leq b_j \leq n_i$ at time $t$ happen on deme $i$ at rate
\begin{equation*}
\frac{ c_N \lambda_i( t )^{ \gamma - 1 } }{ \sum_{ j = 1 }^D \lambda_j( 0 )^{ \gamma + 1 } } \sum_{ l = 0 }^{ ( n_i - b ) \wedge ( 4 - r ) } \binom{ n_i - b }{ l } \frac{ ( 4 )_{ r + l } }{ 4^{ b + l } } \int_0^1 x^{ b + l - 2 } ( 1 - x )^{ n_i - b - l } \Lambda_i( dx ),
\end{equation*}
events in which one lineage branches into $4 L + 1$ lineages occur at rate $n_i \sigma_i / 2$, branching into two lineages due to crossover recombination happens at rate $n_i \rho$, mutations occur ate rate $n_i \theta$, and that migration to deme $j \neq i$ happens at rate $n_i \frac{ \lambda_j( t ) }{ \lambda_i( t ) } m_{ j i }$, where $n_i$ is the number of lineages on deme $i$.
Between migration events, the ancestries of subpopulations on different demes evolve independently.
Convergence will then follow from a straightforward analogue of \cite[Theorem 4.2]{MS2003}.
Throughout, we assume that our sample consists of $n_i$ lineages on deme $i$, and that each lineage carries ancestral material on only one chromosome.
This assumption is justified later by verifying that a separation of timescales phenomenon \citep{M98} takes place, establishing that distinct chromosomes disperse to separate active lineages instantaneously on the coalescent time scale.
\vskip 11pt
\textbf{Multiple mergers via a single large family}

By the Kingman formula for exchangeable, diploid offspring distributions \citep[equation (9)]{MS2003}, the probability of $b \leq n_i$ chromosomes merging by belonging to the same family in the previous time step, and picking the same parental chromosome out of the four possibilities, is
\begin{align*}
&4^{ 1 - b }\frac{ ( M_N^{ ( i ) }( \lfloor t / c_N \rfloor ) )_{ n_i - b + 1 } }{ ( 2 M_N^{ ( i ) }( \lfloor t / c_N \rfloor - 1 ) )_{ n_i } } \E[ ( \nu_1^{ ( i ) } )_b \nu_2^{ ( i ) } \ldots \nu_{ n_i - b + 1 }^{ ( i ) } ].
\end{align*}
Analogously to \citep[equations (28) and (29)]{MS2003}, conditions \eqref{a8} and \eqref{a9} imply that
\begin{align*}
\frac{ M_N^{ ( i ) }( \lfloor t / c_N \rfloor )^{ n_i - b + 1 } }{ 2^{ n_i - b + 1 } M_N^{ ( i ) }( \lfloor t / c_N \rfloor - 1 )^{ n_i } } \E[ ( \nu_1^{ ( i ) } )_b \nu_2^{ ( i ) } \ldots \nu_{ n_i - b + 1 }^{ ( i ) } ] &= c_N^{ ( i ) }( t ) 4^{ 2 - b } \int_0^1 x^{ b - 2 } ( 1 - x )^{ n_i - b } \Lambda_i( dx ) \\
&= \frac{ c_N \lambda_i( t )^{ \gamma - 1 }4^{ 2 - b } }{ \sum_{ j = 1 }^D \lambda_j( 0 )^{ \gamma + 1 } } \int_0^1 x^{ b - 2 } ( 1 - x )^{ n_i - b } \Lambda_i( dx ),
\end{align*}
where the last step follows from \eqref{c_N_equivalence}.
The rate of a particular combination of $r \leq 4$ simultaneous mergers with sizes $b_j \geq 2$ for $j \in [ r ]$ is obtained by summing over all ways in which such a merger can happen, resulting in the overall rate
\begin{equation}\label{xi_rate}
\frac{ c_N \lambda_i( t )^{ \gamma - 1 } }{ \sum_{ j = 1 }^D \lambda_j( 0 )^{ \gamma + 1 } } \sum_{ l = 0 }^{ ( n_i - b ) \wedge ( 4 - r ) } \binom{ n_i - b }{ l } \frac{ ( 4 )_{ r + l } }{ 4^{ b + l } }  \int_0^1 x^{ b + l - 2 } ( 1 - x )^{ n_i - b - l } \Lambda_i( dx ) 
\end{equation}
where $b_1 + \ldots b_r = b \leq n_i$ \cite[equation (27)]{BBE13a}.
\vskip 11pt
\textbf{Multiple mergers via two or more large families}

By \eqref{a8}, the probability of mergers via two or more large families, i.e.~families with at least two offspring in the sample, is bounded from above by 
\begin{align*}
\frac{ 1 }{ ( 2 M_N^{ ( i ) }( \lfloor t / c_N \rfloor - 1 ) )_{ n_i } } \sum_{ j_1 \neq j_2 = 1 }^{ M_N^{ ( i ) }( \lfloor t / c_N \rfloor ) } &\E\left[ ( \nu_{ j_1 }^{ ( i ) } + \tilde{ \beta }_{ j_1 }^{ ( i ) } )_2 ( \nu_{ j_2 }^{ ( i ) } + \tilde{ \beta }_{ j_2 }^{ ( i ) } )_2 \left( \sum_{ k = 1 }^{ M_N^{ ( i ) }( \lfloor t / c_N \rfloor ) } \nu_k^{ ( i ) } + \beta_k^{ ( i ) } \right)^{ n_i - 4 } \right] \\
&\leq \frac{ ( M_N^{ ( i ) }( \lfloor t / c_N \rfloor ) )_2 }{ ( 2 M_N^{ ( i ) }( \lfloor t / c_N \rfloor - 1 ) )_4 } \E[ ( \nu_1^{ ( i ) } + \tilde{ \beta }_1^{ ( i ) } )_2 ( \nu_j^{ ( i ) } + \tilde{ \beta }_2^{ ( i ) } )_2 ] = o( c_N ).
\end{align*}
\vskip 11pt
\textbf{A single migration event}

The event that all $n_i$ lineages belong to different families in the previous generation, and that one individual migrates from deme $i$ to $j$ in reverse time, has asymptotic probability
\begin{align*}
&\left( 1 - \sum_{ k = 1 }^D m_{ i k }^{ ( N ) } \right)^{ n_i - 1 } n_{ i } \frac{ M_N^{ ( j ) }( \lfloor t / c_N \rfloor - 1 ) m_{ j i }^{ ( N ) } }{ M_N^{ ( i ) }( \lfloor t / c_N \rfloor - 1 ) } \sim c_N n_i \frac{ \lambda_j( t ) }{ \lambda_i( t ) } m_{ j i },
\end{align*}
by \eqref{reversible_migration}, \eqref{a7}, and \eqref{a12}.

A similar calculation demonstrates that the analogous probability for more than one simultaneous migration event is $o( c_N )$, while combining the above with the first two calculations demonstrates that a single migration occurring simultaneously with one or more large families is also an $o( c_N )$ event.
\vskip 11pt
\textbf{A single mutation event}

An analogous calculation to the migration case using \eqref{a10} shows that the probability of one site mutating in the previous time step with no other accompanying events converges to $c_N n_i \theta$.
\vskip 11pt
\textbf{A single recombination event}

Likewise, an analogous calculation to the migration case using \eqref{a11} shows that the probability of one chromosome recombining in the previous time step with no other accompanying events converges to $c_N n_i \rho$.
\vskip 11pt
\textbf{A single branching event due to a selective birth}

The probability of a lineage belonging to a selective birth by a family in the previous generation depends on the fitness of its parents, which is unknown.
An elegant solution is to add selective events at the greatest possible rate, add the $4 L$ chromosomes belonging to the two potential parents into the sample along with retaining the child lineage whenever a selection event happens, and track this extended sample to its \emph{ultimate ancestor}: the most recent common ancestor of the original sample, as well as all potential selective parents encountered along the way \citep{KN97, NK97}.
The type of the ultimate ancestor can then be sampled from the stationary distribution of $M$ (or any other desired initial law), with mutations occurring along lineages and alleles propagated to children as before.
Now the alleles, and thus the fitness of the selective parents are known at each potential selective event, and the true ancestry of each child lineage can be assigned to either a randomly chosen parent with probability $\E[ \beta_j^{ ( i ) } ] / \E[ \tilde{ \beta }_j^{ ( i ) } ]$, or to remain with the ongoing child lineage with the complementary probability.

From the point of view of the ancestral process, such selective branching events in which one lineage on deme $i \in [ D ]$ branches into $4 L + 1$ lineages (corresponding to the single-chromosome child lineage, as well as the $4 L$ parental chromosomes which immediately disperse into separate lineages due to separation of time scales) happens with asymptotic probability
\begin{align*}
&\frac{ 1 }{ ( 2 M_N^{ ( i ) }( \lfloor t / c_N \rfloor - 1 ) )_{ n_i } } \sum_{ \substack{ j_1 \neq \ldots \neq j_{ n_i } = 1 \\ \text{all distinct} } }^{ M_N^{ ( i ) }( \lfloor t / c_N \rfloor ) } \E[ \tilde{ \beta }_{ j_1 }^{ ( i ) } \nu_{ j_2 }^{ ( i ) } \ldots \nu_{ j_{ n_i } }^{ ( i ) }  ] \\
&= \frac{ 1 }{ ( 2 M_N^{ ( i ) }( \lfloor t / c_N \rfloor - 1 ) )_{ n_i } } \sum_{ j = 1 }^{ M_N^{ ( i ) }( \lfloor t / c_N \rfloor ) } \E\left[ \tilde{ \beta }_j^{ ( i ) } \left( \sum_{ k \neq j }^{ M_N^{ ( i ) }( \lfloor t / c_N \rfloor ) } \nu_k^{ ( i ) } \right)^{ n_i - 1 } \right]  \\
&= \frac{ M_N^{ ( i ) }( \lfloor t / c_N \rfloor ) }{ ( 2 M_N^{ ( i ) }( \lfloor t / c_N \rfloor - 1 ) )_{ n_i } } \E\left[ \tilde{ \beta }_1^{ ( i ) } \left( 2 M_N^{ ( i ) }( \lfloor t / c_N \rfloor - 1 ) - \nu_1^{ ( i ) } - \sum_{ k = 1 }^{ M_N^{ ( i ) }( \lfloor t / c_N \rfloor ) } \beta_k^{ ( i ) } \right)^{ n_i - 1 } \right].
\end{align*}
A binomial expansion followed by \eqref{a13} and \eqref{a14} yield
\begin{align*}
&\frac{ 1 }{ ( 2 M_N^{ ( i ) }( \lfloor t / c_N \rfloor - 1 ) )_{ n_i } } \sum_{ \substack{ j_1 \neq \ldots \neq j_{ n_i } = 1 \\ \text{all distinct} } }^{ M_N^{ ( i ) }( \lfloor t / c_N \rfloor ) } \E[ \tilde{ \beta }_{ j_1 }^{ ( i ) } \nu_{ j_2 }^{ ( i ) } \ldots \nu_{ j_{ n_i } }^{ ( i ) }  ] \\
&= \frac{ M_N^{ ( i ) }( \lfloor t / c_N \rfloor ) }{ ( 2 M_N^{ ( i ) }( \lfloor t / c_N \rfloor - 1 ) )_{ n_i } } \sum_{ l = 0 }^{ n_i - 1 } \binom{ n_i - 1 }{ l } [ 2 M_N^{ ( i ) }( \lfloor t / c_N \rfloor - 1 ) ]^{ n_i - 1 - l } ( -1 )^l \\
&\phantom{ = } \times \E\left[ \tilde{ \beta }_1^{ ( i ) } \left( \nu_1^{ ( i ) } + \sum_{ k = 1 }^{ M_N^{ ( i ) }( \lfloor t / c_N \rfloor ) / 2 } \beta_k^{ ( i ) } \right)^l \right] \\
&\sim \frac{ M_N^{ ( i ) }( \lfloor t / c_N \rfloor ) }{ 2 M_N^{ ( i ) }( \lfloor t / c_N \rfloor - 1 ) } \E[ \tilde{ \beta }_1^{ ( i ) } ] + o( c_N ) \sim c_N \frac{ \sigma_i }{ 2 } + o( c_N ),
\end{align*}
as required.
\vskip 11pt
\textbf{Multiple simultaneous branching events}

Multiple simultaneous selective events can take place in one of three ways: two (or more) simultaneous selective births in the same family, two (or more) simultaneous selective births in a combination of families, or a combination of selective and neutral births in the same family.
The probability of all three kinds of events is bounded above by

\begin{align*}
&\frac{ M_N^{ ( i ) }( \lfloor t / c_N \rfloor ) }{ ( 2 M_N^{ ( i ) }( \lfloor t / c_N \rfloor - 1 ) )_{ n_i } } \E\left[ \tilde{ \beta }_1^{ ( i ) } \left\{ 2 M_N^{ ( i ) }( \lfloor t / c_N \rfloor )^{ n_i - 1 } - \sum_{ \substack{ i_2 \neq \ldots \neq i_a \neq 1 \\ \text{all distinct} } }^{ M_N^{ ( i ) }( \lfloor t / c_N \rfloor ) } \nu_{ i_2 }^{ ( i ) } \ldots \nu_{ i_a }^{ ( i ) } \right\} \right] \\
&\leq \frac{ M_N^{ ( i ) }( \lfloor t / c_N \rfloor ) }{ ( 2 M_N^{ ( i ) }( \lfloor t / c_N \rfloor - 1 ) )_{ n_i } } \E\Bigg[ \tilde{ \beta }_1^{ ( i ) } \Bigg\{ 2 M_N^{ ( i ) }( \lfloor t / c_N \rfloor - 1 )^{ n_i - 1 } \\ 
&\phantom{= \frac{ M_N^{ ( i ) }( \lfloor t / c_N \rfloor ) }{ 2 ( M_N^{ ( i ) }( \lfloor t / c_N \rfloor - 1 ) )_{ n_i } } \E}- \left( 2 M_N^{ ( i ) }( \lfloor t / c_N \rfloor - 1 ) - \sum_{ j = 1 }^{ M_N^{ ( i ) }( \lfloor t / c_N \rfloor ) } \tilde{ \beta }_j^{ ( i ) } \right)^{ n_i - 1 } \Bigg\} \Bigg] \\
&= o( c_N ),
\end{align*}
by a binomial expansion and \eqref{a14}.
\vskip 11pt
\textbf{Dispersal of chromosomes into distinct, single-marked individuals}

Finally, we abandon the assumption that all lineages carry ancestral material on only one chromosome in order to verify the separation of time scales phenomenon.
The probability that $n / 2$ individuals with ancestral material both chromosomes (or so-called \emph{double-marked} individuals) in a pair disperse into $n$ parents, each of whom carries ancestral material on only one copy of the chromosome (so-called \emph{single-marked} individuals), in the previous generation is $O( 1 )$. To see why, note that every individual is replaced at every time step, and individuals always inherit one chromosome from each parent.
Thus, complete dispersal of $n / 2$ double-marked individuals happens in one generation provided that all $n / 2$ individuals originate from different families, which has probability at least
\begin{equation*}
\prod_{ i = 1 }^D \frac{ ( M_N^{ ( i ) }( t ) )_{ n_i } }{ ( 2 M_N^{ ( i ) }( t - 1 ) )_{ n_i } } \E[ \nu_1^{ ( i ) } \ldots \nu_{ n_i }^{ ( i ) } ] = O( 1 ).
\end{equation*}
Likewise, the probability of two active chromosomes splitting apart into distinct ancestors is $1 / 2 = O ( 1 )$ because assignments of parents to chromosomes is done independently and uniformly at random.
Hence, the probability of a lineage with $2 L$ ancestral chromosomes dispersing into $2 L$ lineages with a single ancestral chromosome each in at most $2 L - 1$ generations happens with probability at least
\begin{equation*}
\left( \frac{ 1 }{ 2 } \right)^{ 2 L }\prod_{ t = 1 }^{ 2 L } \prod_{ i = 1 }^D \frac{ ( M_N^{ ( i ) }( t ) )_{ n_i } }{ ( 2 M_N^{ ( i ) }( t - 1 ) )_{ n_i } } \E[ \nu_1^{ ( i ) } \ldots \nu_{ n_i }^{ ( i ) } ] = O( 1 ).
\end{equation*}
Probabilities of merger, recombination, selection or migration events were all established above to be $O( c_N )$, and thus the probability of complete dispersal before any merger, recombination, selection or migration events is of order
\begin{equation*}
\frac{ 1 }{ 1 + A c_N } \rightarrow 1,
\end{equation*}
where $A > 0$ is a constant independent of both $n$ and $N$.
Thus an analogue of the separation of timescales result in \citep{M98} holds, which justifies considering only single-marked configurations in the previous computations of transition probabilities.

\section{Robustness results}\label{robustness}

The following three subsections quantify the respective effect of selection, recombination, and population structure on the sampling distribution of the singleton-tail statistic.
Each subsection specialises the model of Section \ref{alt_models} to consist of only the relevant force by a particular choice of parameters.
We assume the model of \cite{S2003} for the evolution of the population, and thus consider a one-dimensional family of coalescents specified by $\Lambda_i( dx ) = \operatorname{Beta}( 2 - \alpha, \alpha )( dx )$ in \eqref{a9} for $\alpha \in ( 1, 2 )$, with corresponding time scaling $c_N \sim N^{ 1 - \alpha }$ and $\gamma = 2 - \alpha$ in \eqref{a6}.
Under the alternative hypothesis $\alpha < 2$, the population sizes on demes will be constant, i.e.~$\lambda_i( t ) = d_i$ for relative deme sizes $d_1 + \ldots + d_D = 1$. Under the null hypothesis $\alpha = 2$, populations on demes will undergo exponential growth forwards in time, corresponding to $M_N^{ ( i ) }( t ) := \lfloor N d_i ( 1 + \gamma_N )^{ -t } \rfloor$, resulting in $\lambda_i( t ) = d_i e^{ - \gamma t }$ for the population-rescaled growth rate $\gamma = \lim_{ N \rightarrow \infty } \gamma_N / c_N$.

It will also be necessary to distinguish between two kinds of data sets: simulated data used to fit KDEs to approximate likelihoods, and compute the quantile $q_{ \omega }$ in \eqref{test}, as well as observed data, which will also be simulated in this instance, but which will typically be a biological data set.
We will refer to the former as calibration data, and the latter as pseudo-observed data.
Pseudo-observed data is reserved solely for plugging into KDE approximations of likelihoods (computed from calibration data) to obtain likelihood ratio test statistics.
A C++ implementation of the algorithm used to generate the data in this section is available at \href{https://github.com/JereKoskela/Beta-Xi-Sim}{https://github.com/JereKoskela/Beta-Xi-Sim}.

We set the number of simulation replicates per model at 1000 (note that $\Theta_0$ contains 133 models, and $\Theta_1$ a further 41), the sample size at $n = 500$, the lumped tail cutoff at $k = 15$, and assume the true mutation rate is known.
The number of unlinked chromosomes per sample is set to 23 to match the number of linkage groups in Atlantic cod \citep[Supplementary Table 3]{T17} --- an organism for which multiple merger have frequently been suggested as an important evolutionary mechanism \citep{SBB2013, TL14}.
Results are averaged across chromosomes as outlined in Section \ref{single_tail}.
The lengths of the 23 chromosomes have also been set (by multiplying the total rate of mutation on each chromosome by the number of sites it contains) to match those reported in \citep[Supplementary Table 3]{T17}.
The approximate size of hypothesis tests is set at $\omega = 0.01$ throughout.

\subsection{Weak selection}

In this section we consider the model of Section \ref{alt_models} with a single deme ($D = 1$), and no recombination ($\rho = 0$).
The resulting process is a $\Xi$-coalescent analogue of the Complex Selection Graph (CSG) \cite{F03}.
We compute realisations of the singleton-tail statistic by assuming that neutral, infinitely-many-sites mutations occur on each chromosome along the branches of the realised non-neutral tree sampled from the $\Xi$-CSG, but that the selective types of individuals are unobserved.
This assumption is reasonable if the fitness of individuals cannot be observed, or if mutations with a fitness effect are either much less frequent than neutral mutations, or occur in unobserved regions.

\begin{figure}[!ht]
\centering
\includegraphics[width = 0.49 \linewidth]{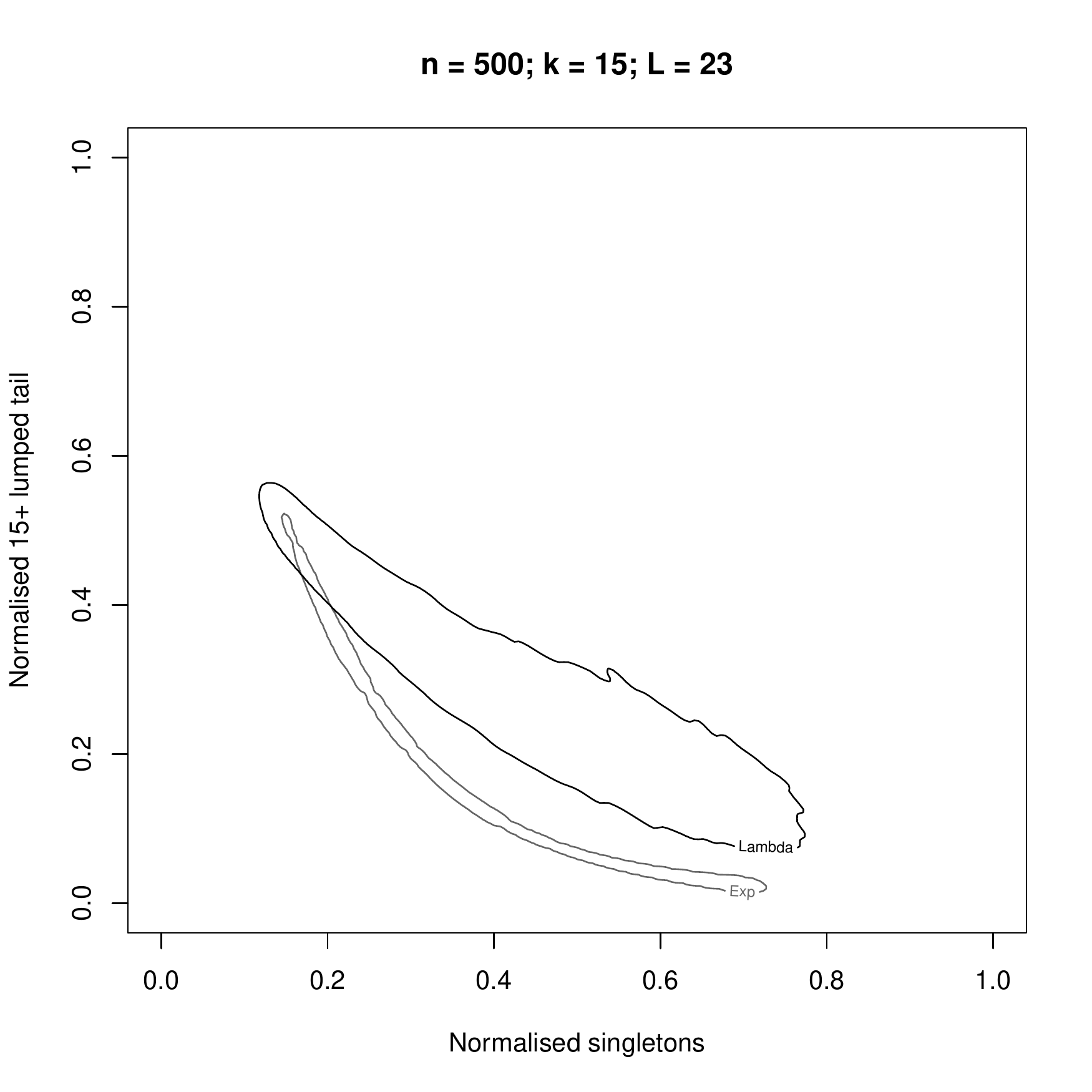}
\includegraphics[width = 0.49 \linewidth]{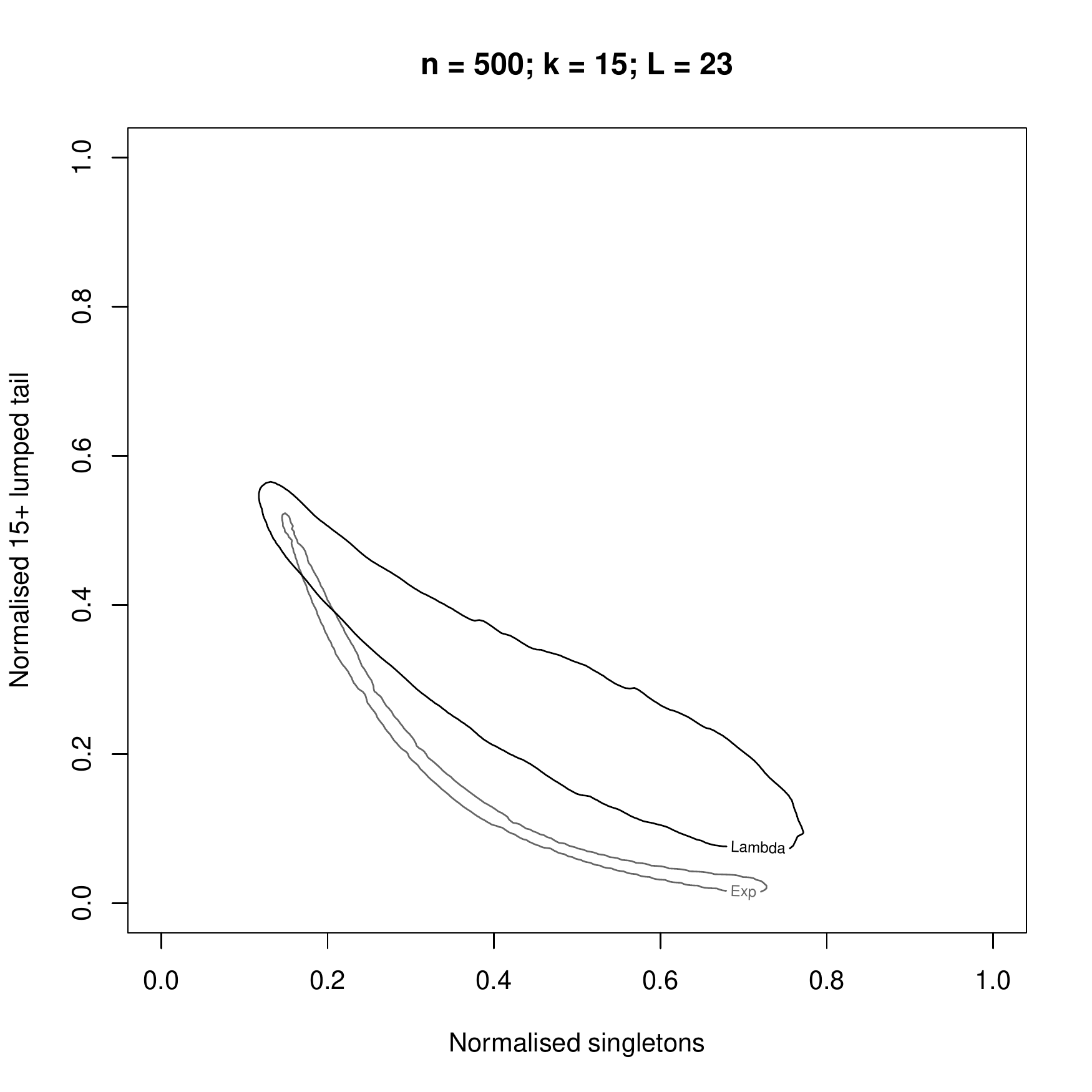}
\caption{$99^{\text{th}}$ percentiles of KDEs fitted to 1000 realisations of the singleton-tail statistic for each model in $\Theta_0$ and $\Theta_1$.
Each sample consists of 23 chromosomes, and (Left) $\sigma = 0$, or (Right) $\sigma \in (0.0000016, 0.0008)$ per chromosome pair as $\alpha$ varies from 1 to 2.}
\label{selection}
\end{figure}

Figure \ref{selection} shows the sampling distributions of the neutral and non-neutral models.
The fitness model assumes two alleles, a and A, with each chromosome pair contributing fitness $\sigma > 0$ if either parent carries at least one A allele at that pair, and 0 otherwise.
The selection rates are necessarily low, because the cost of simulating the ASGs is known to increase exponentially in $\sigma$ \citep[Appendix A]{F01}.
Efficiency gains resulting from perfect simulation techniques \citep{F01, F03} cannot be employed because they rely on terminating the simulation before reaching the MRCA, and thus the SFS cannot be resolved.

\begin{figure}[!ht]
\centering
\includegraphics[width = 0.49 \linewidth]{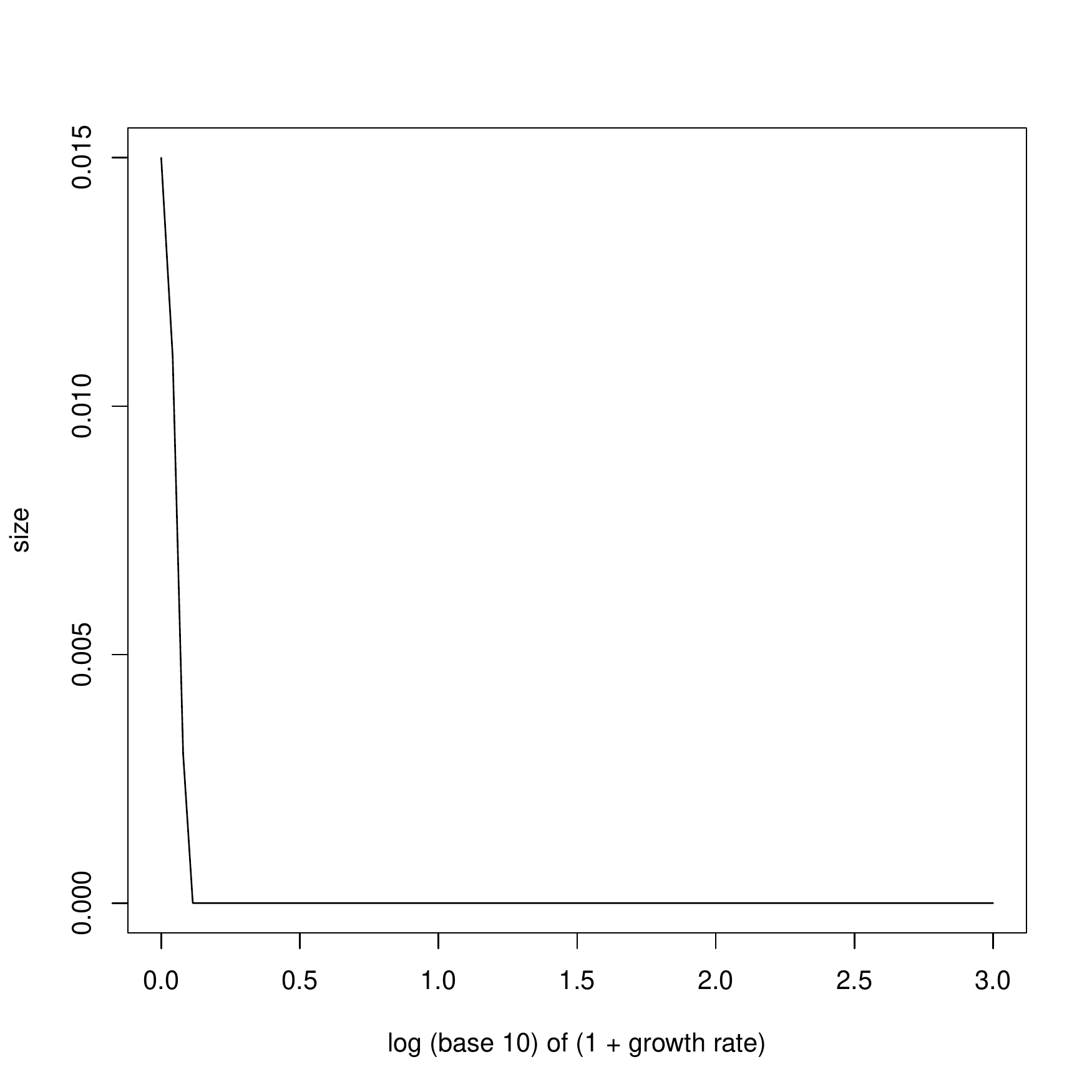}
\includegraphics[width = 0.49 \linewidth]{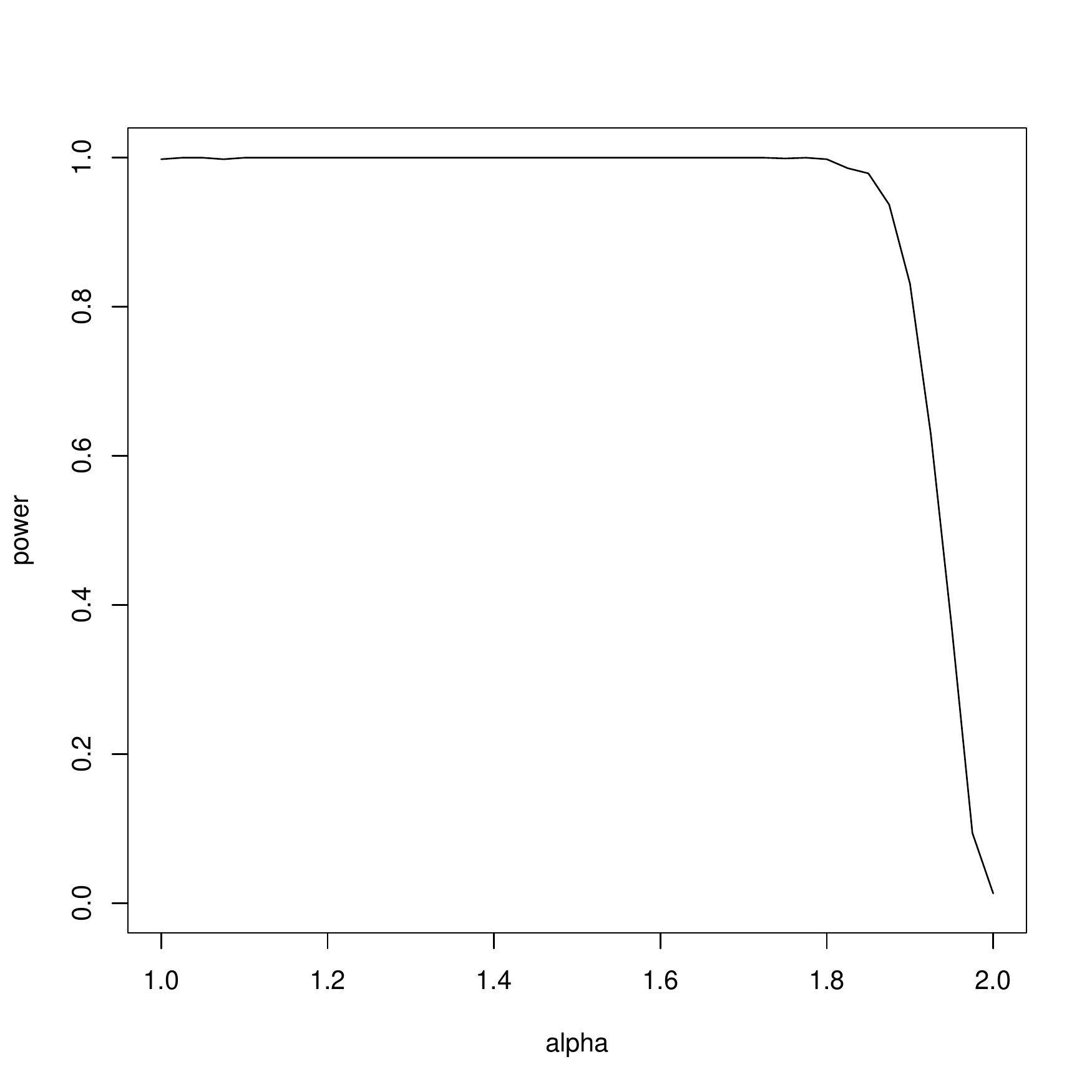}
\caption{Empirical size (Left) and power (Right) of a $\Theta_0$ vs $\Theta_1$ test conducted using calibration data simulated under neutral models, but applied to pseudo-observed data simulated from selective models. The simulation parameters are as in Figure \ref{selection}.}
\label{test_selection}
\end{figure}

The results in Figure \ref{selection} show striking agreement between sampling distributions in the neutral and selective cases.
We also conducted the hypothesis test \eqref{test} using calibration data simulated from a neutral model, and applied the resulting misspecified test to pseudo-observed data simulated from a model with weak selection.
Figure \ref{test_selection} shows that the performance of the test was excellent, with high power and size well below the formal threshold of $\omega = 0.01$ for the majority of the parameter ranges.

\begin{figure}[!ht]
\centering
\includegraphics[width = 0.49 \linewidth]{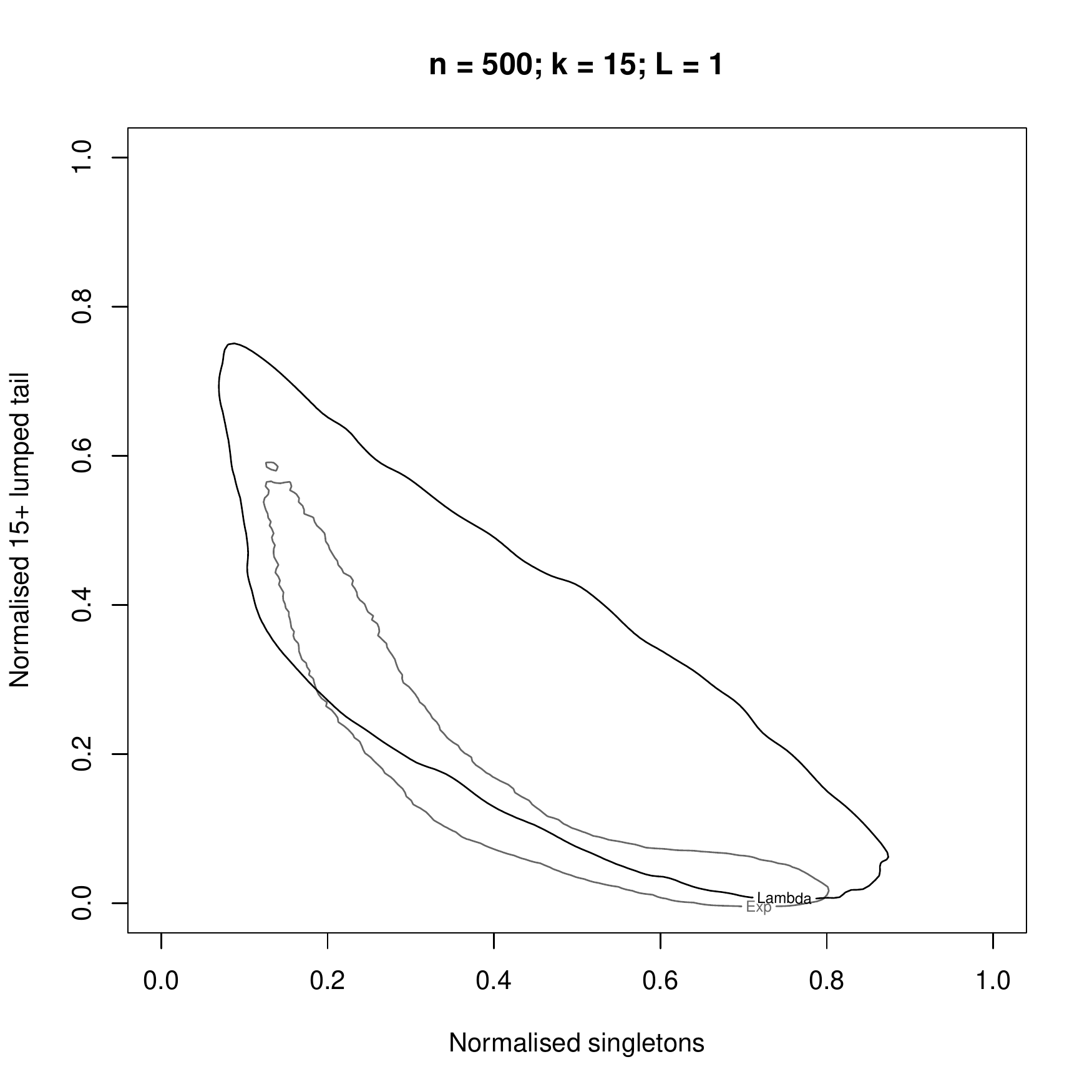}
\includegraphics[width = 0.49 \linewidth]{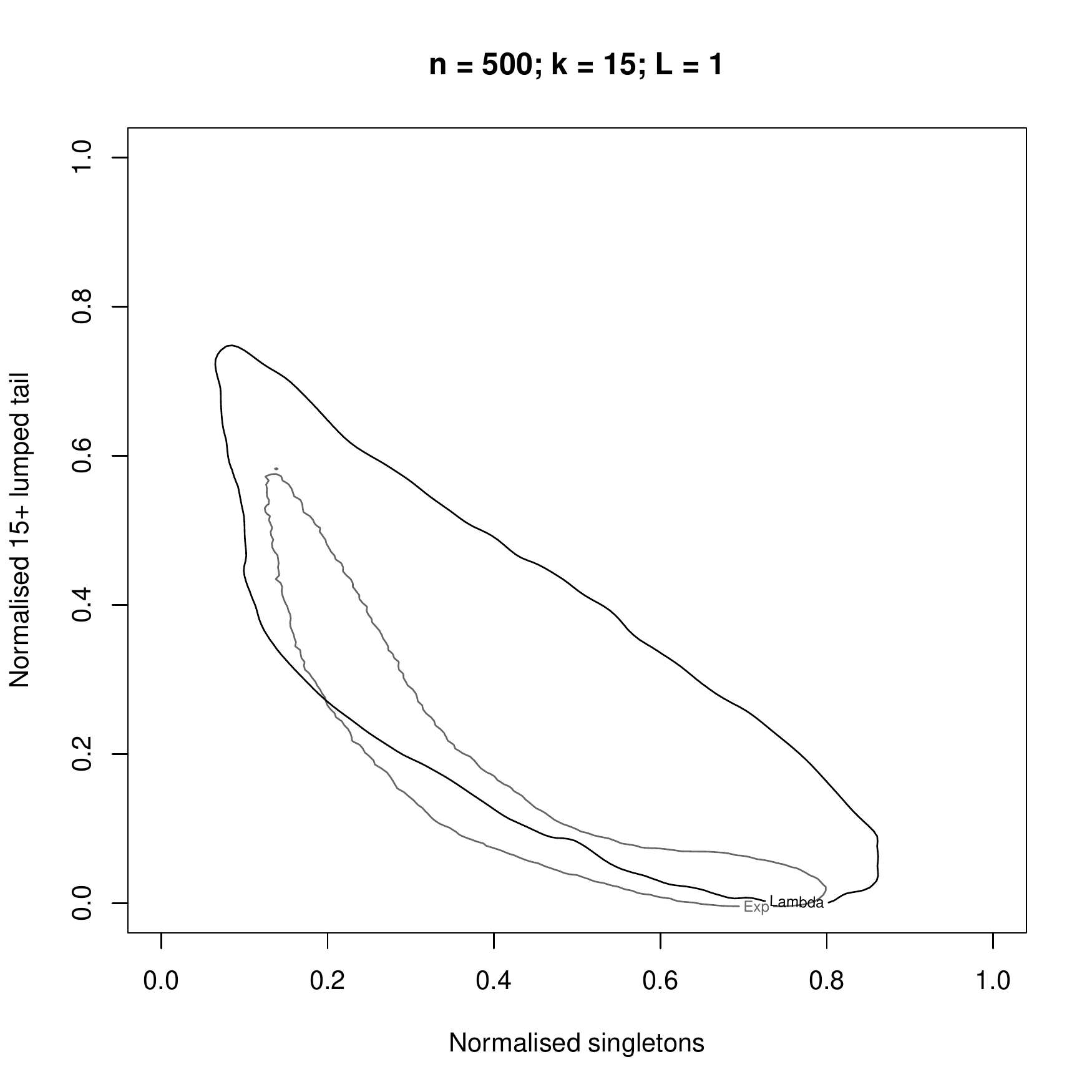}
\caption{$99^{\text{th}}$ percentiles of KDEs fitted to 1000 realisations of the singleton-tail statistic for each model in $\Theta_0$ and $\Theta_1$.
Each sample consists of one chromosome, and (Left) $\sigma = 0$, or (Right) $\sigma \in (0.0024, 1.2)$ as $\alpha$ varies from 1 to 2.}
\label{single_locus_selection}
\end{figure}

To investigate the effect of a larger selection coefficient, we also simulated realisations of the singleton-tail statistic under a single chromosome model.
In this setting, each selective branching event results in five lineages, as opposed to $4 \times 23 + 1 = 93$ as in the 23 chromosome case.
The sampling variance under a single locus is too large for a powerful statistical test, but Figure \ref{single_locus_selection} demonstrates that the sampling distributions with and without selection remain very similar.
Taken together, these simulations show that the distribution of realised relative branch lengths under the CSG is similar to that under a neutral coalescent, at least for external branches, as well as for the oldest branches before the MRCA is reached.
Hence, the singleton-tail statistic cannot be used to detect weak selection, but can discriminate between population growth and $\Xi$-coalescents without knowing whether weak selection is taking place.

\subsection{Recombination}

In this section we consider the model of Section \ref{alt_models} with a single deme ($D = 1$), and no selection ($\sigma_i \equiv 0$).
Realisations of the singleton-tail statistic are computed by assuming a neutral, infinitely-many-sites mutation model along the branches of the realised $\Xi$-Ancestral Recombination Graph.

\begin{figure}[!ht]
\centering
\includegraphics[width = 0.49 \linewidth]{scatter_neutral.pdf}
\includegraphics[width = 0.49 \linewidth]{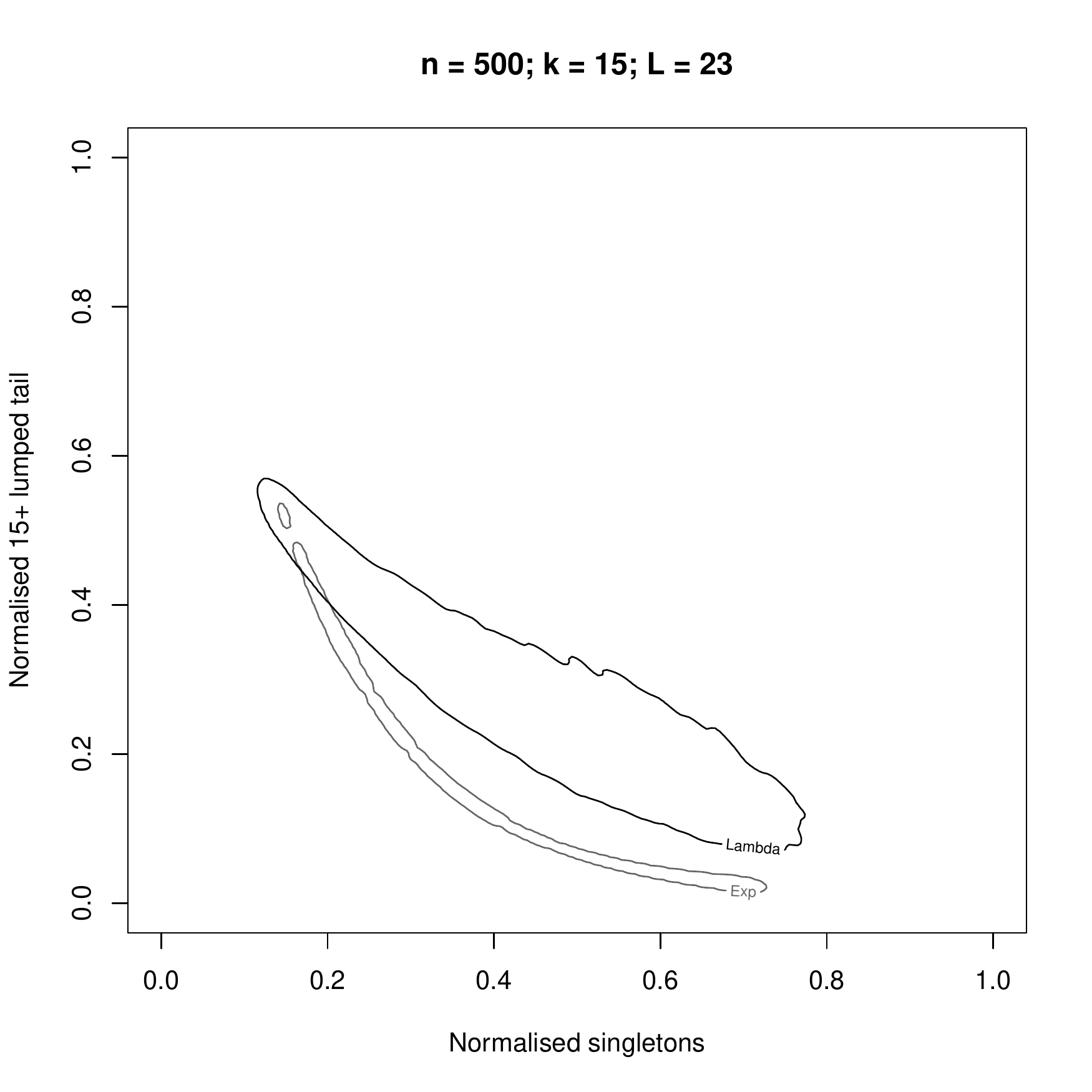}
\caption{$99^{\text{th}}$ percentiles of KDEs fitted to 1000 realisations of the singleton-tail statistic for each model in $\Theta_0$ and $\Theta_1$.
(Left) $\rho = 0$.
(Right) $\rho \in ( 0.001, 0.5 )$ as $\alpha$ varies from 1 to 2.}
\label{recombination}
\end{figure}

\begin{figure}[!ht]
\centering
\includegraphics[width = 0.49 \linewidth]{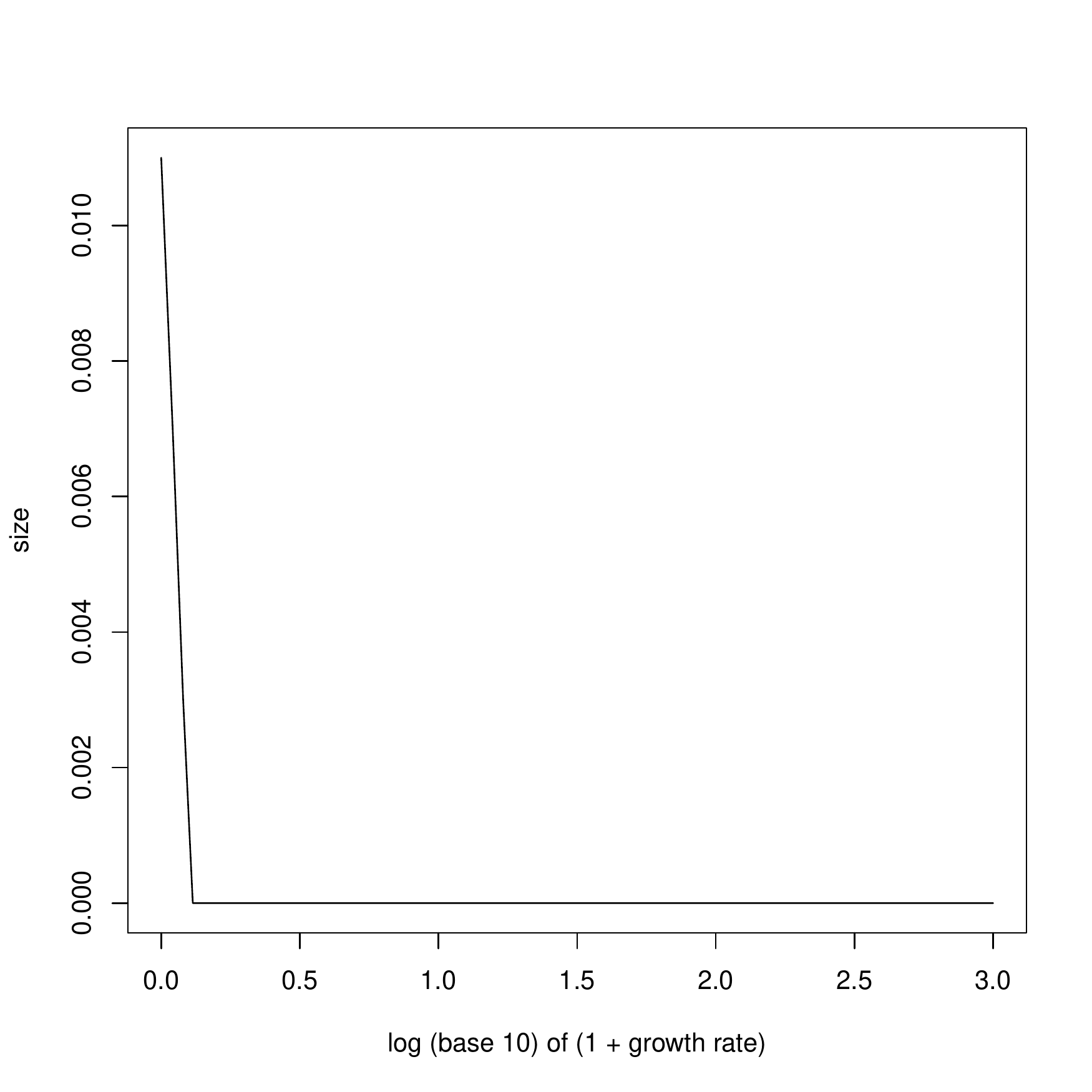}
\includegraphics[width = 0.49 \linewidth]{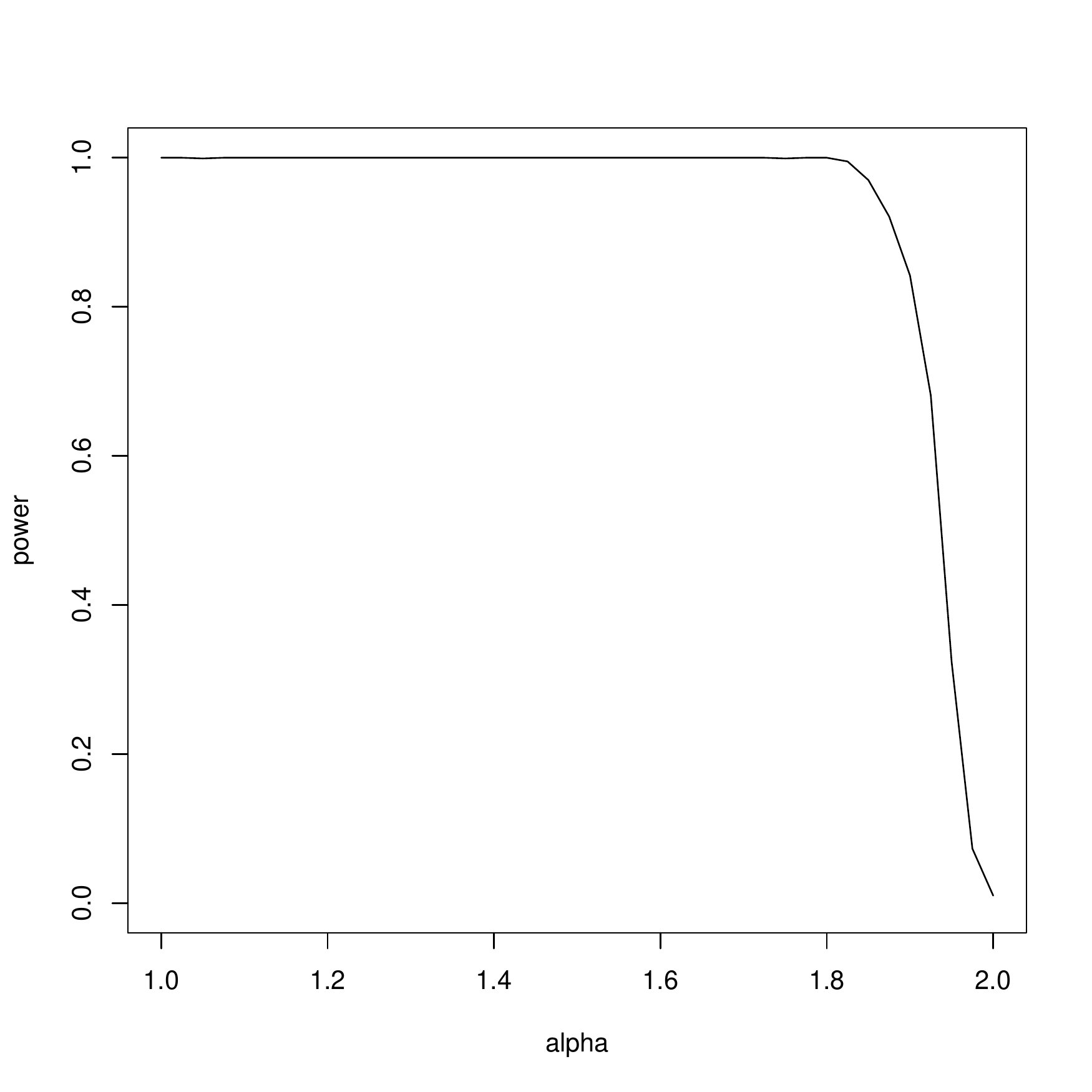}
\caption{Empirical size (Left) and power (Right) of a $\Theta_0$ vs $\Theta_1$ test conducted using calibration data from models without recombination, but applied to pseudo-observed data from models with recombination. The simulation parameters are as in Figure \ref{recombination}.}
\label{test_recombination}
\end{figure}

Figure \ref{recombination} presents a comparison between models with and without recombination.
As was the case with weak selection (see Figure \ref{selection}), the presence of recombination makes no discernible difference to the sampling distribution of the singleton-tail statistic, although the distribution of intermediate SFS entries was observed to be different (results not shown).
Figure \ref{test_recombination} demonstrates that the size and power of statistical tests are unaffected when a misspecified model which wrongly neglects recombination is used to generate calibration data, and the hypothesis test is conducted on pseudo-observed data with recombination.

\subsection{Population structure}

In this section we consider the model of Section \ref{alt_models} with a no selection ($\sigma_i \equiv 0$), no recombination ($\rho = 0$), and two different patterns of population structure.

\begin{figure}[!ht]
\centering
\includegraphics[width = 0.49 \linewidth]{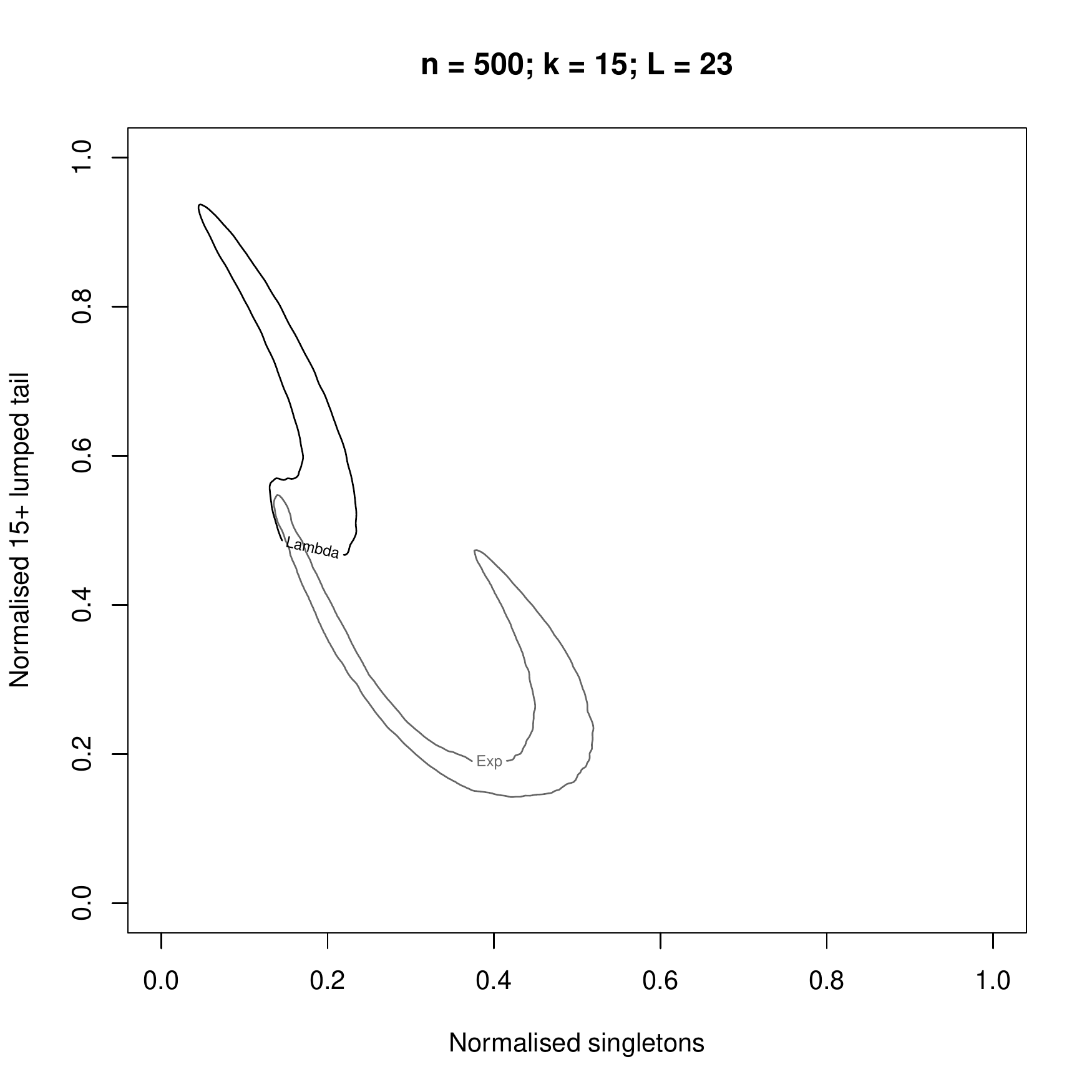}
\includegraphics[width = 0.49 \linewidth]{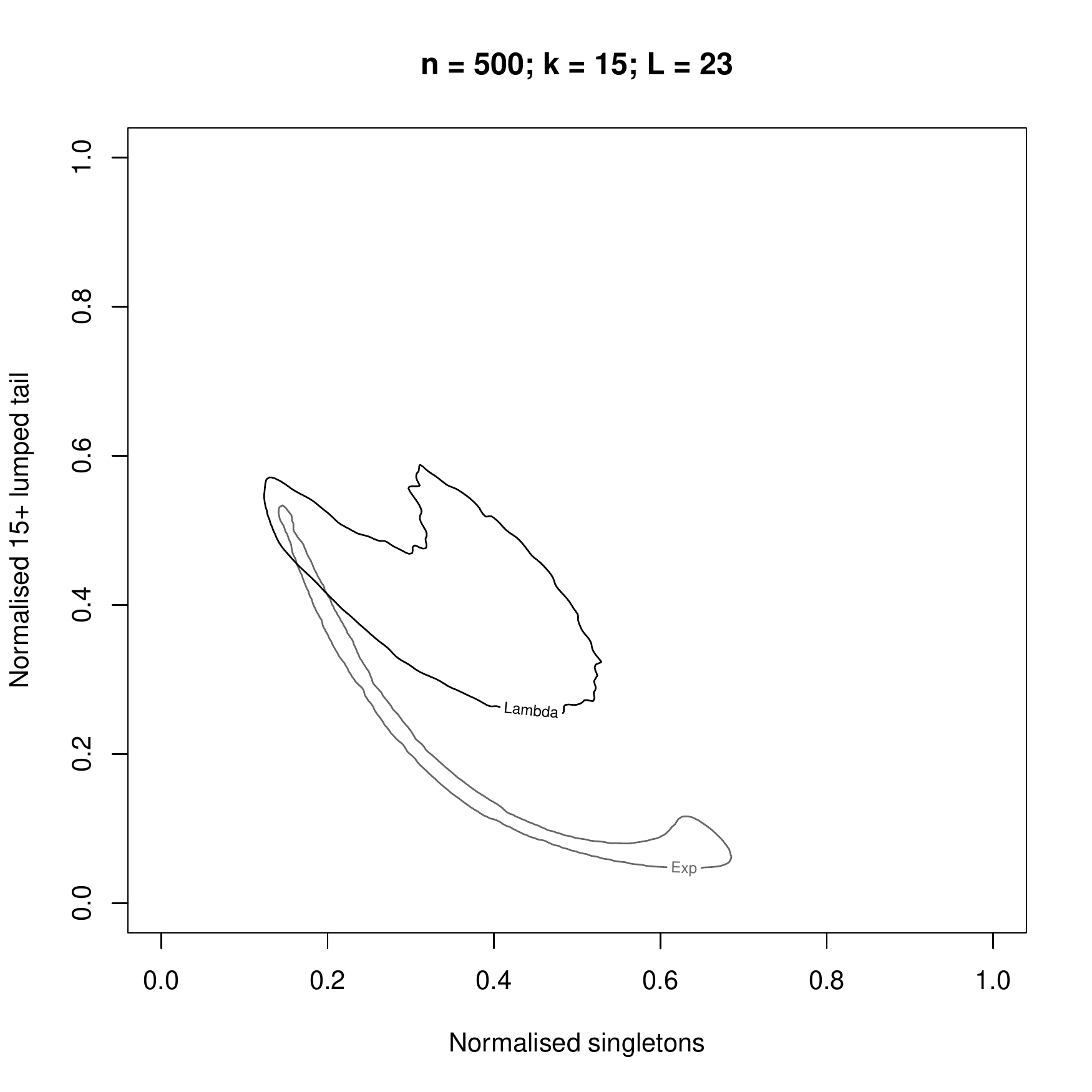}
\caption{$99^{\text{th}}$ percentiles of KDEs fitted to 1000 realisations of the singleton-tail statistic for each model in $\Theta_0$ and $\Theta_1$.
(Left) Four demes with equal population sizes and symmetric migration between all pairs of demes at reverse-time rate $\tilde{ m } \in ( 0.01, 5 )$ as $\alpha$ varies from 1 to 2.
(Right) Two demes with relative population sizes $( 0.75, 0.25 )$, and reverse-time migration rates ranging from $( \tilde{ m }_{ 1 2 }, \tilde{ m }_{ 2 1 } ) = ( 0.01, 0.03 )$ when $\alpha = 1$ to $( \tilde{ m }_{ 1 2 }, \tilde{ m }_{ 2 1 } ) = ( 5, 15 )$ when $\alpha = 2$.}
\label{structure}
\end{figure}

\begin{figure}[!ht]
\centering
\includegraphics[width = 0.49 \linewidth]{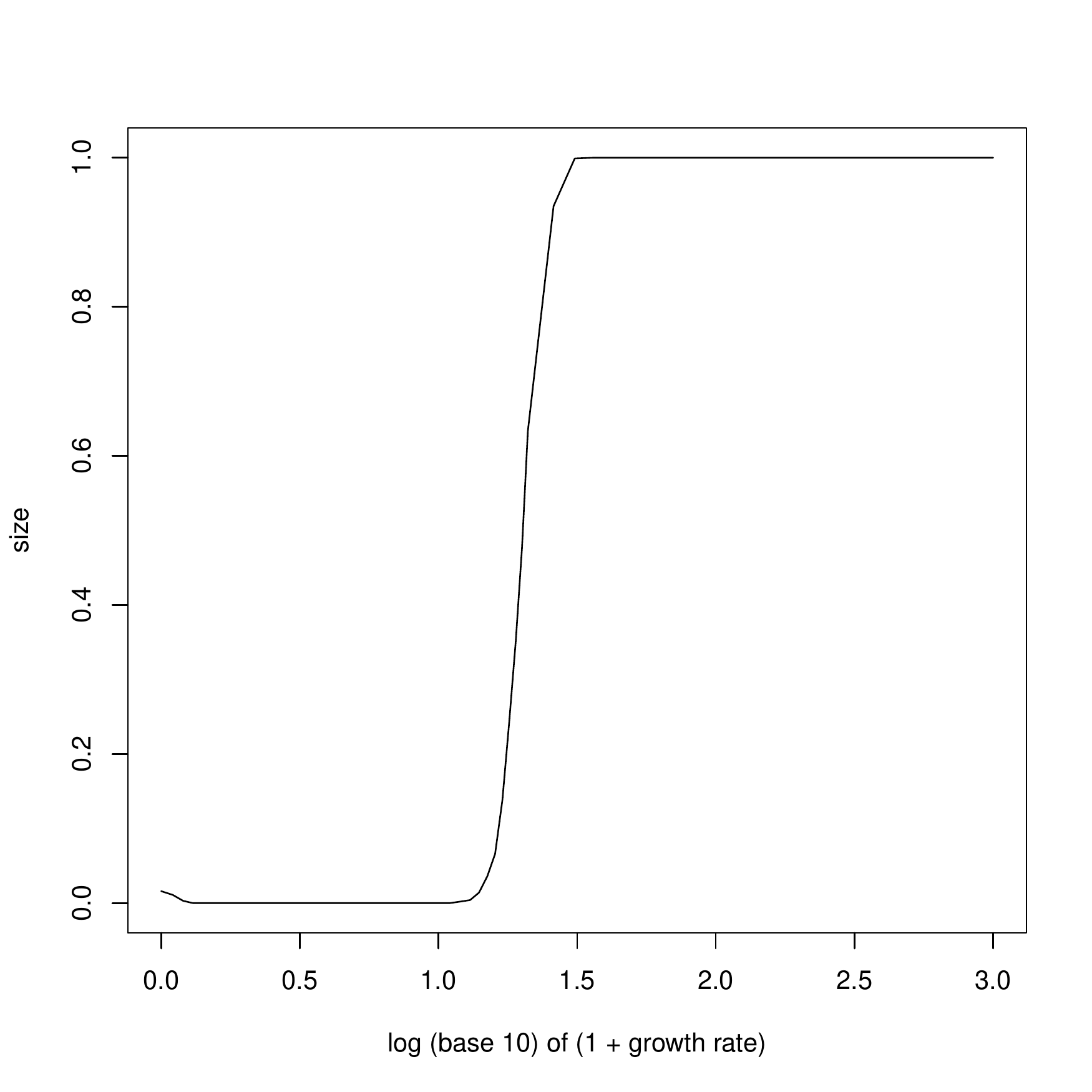}
\includegraphics[width = 0.49 \linewidth]{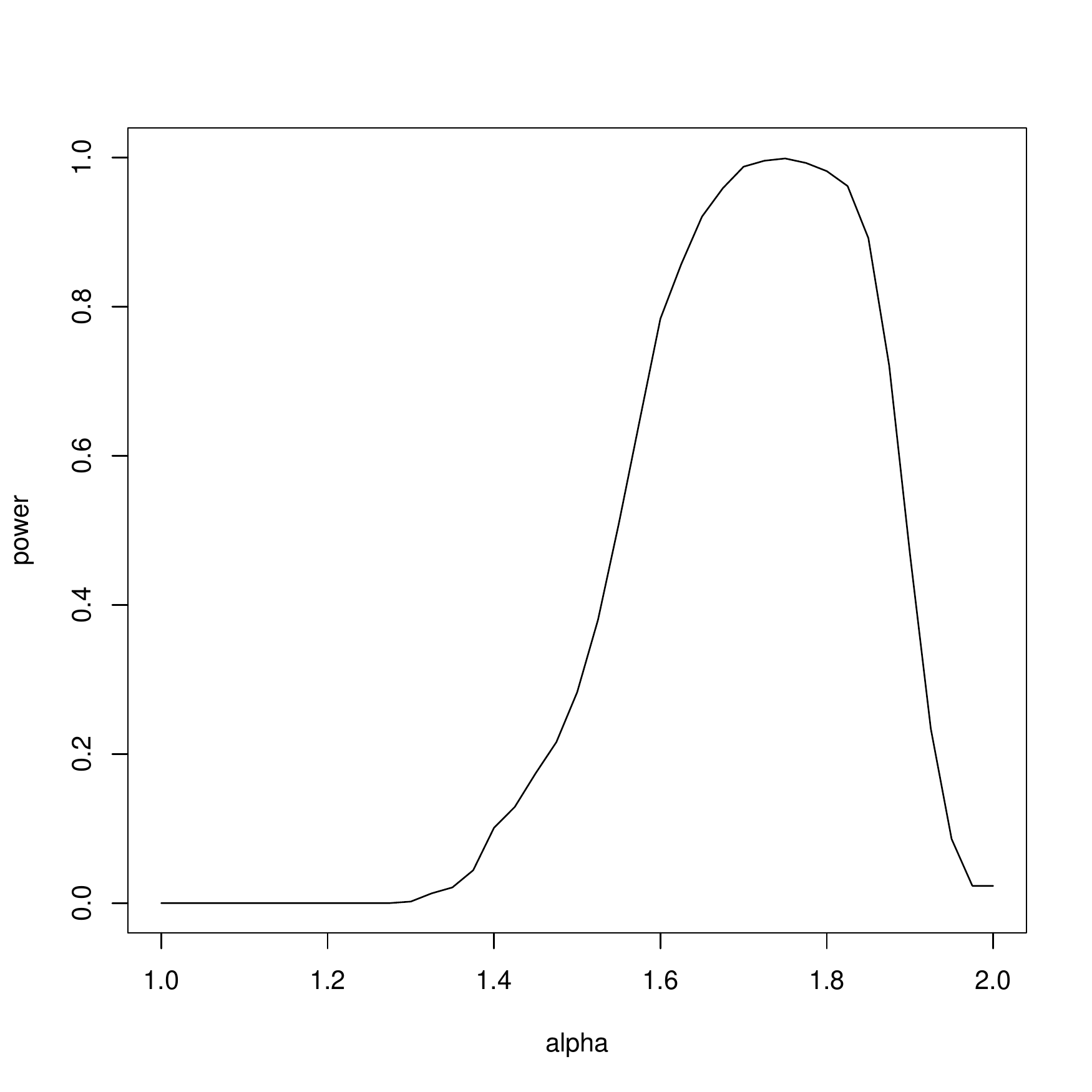}
\includegraphics[width = 0.49 \linewidth]{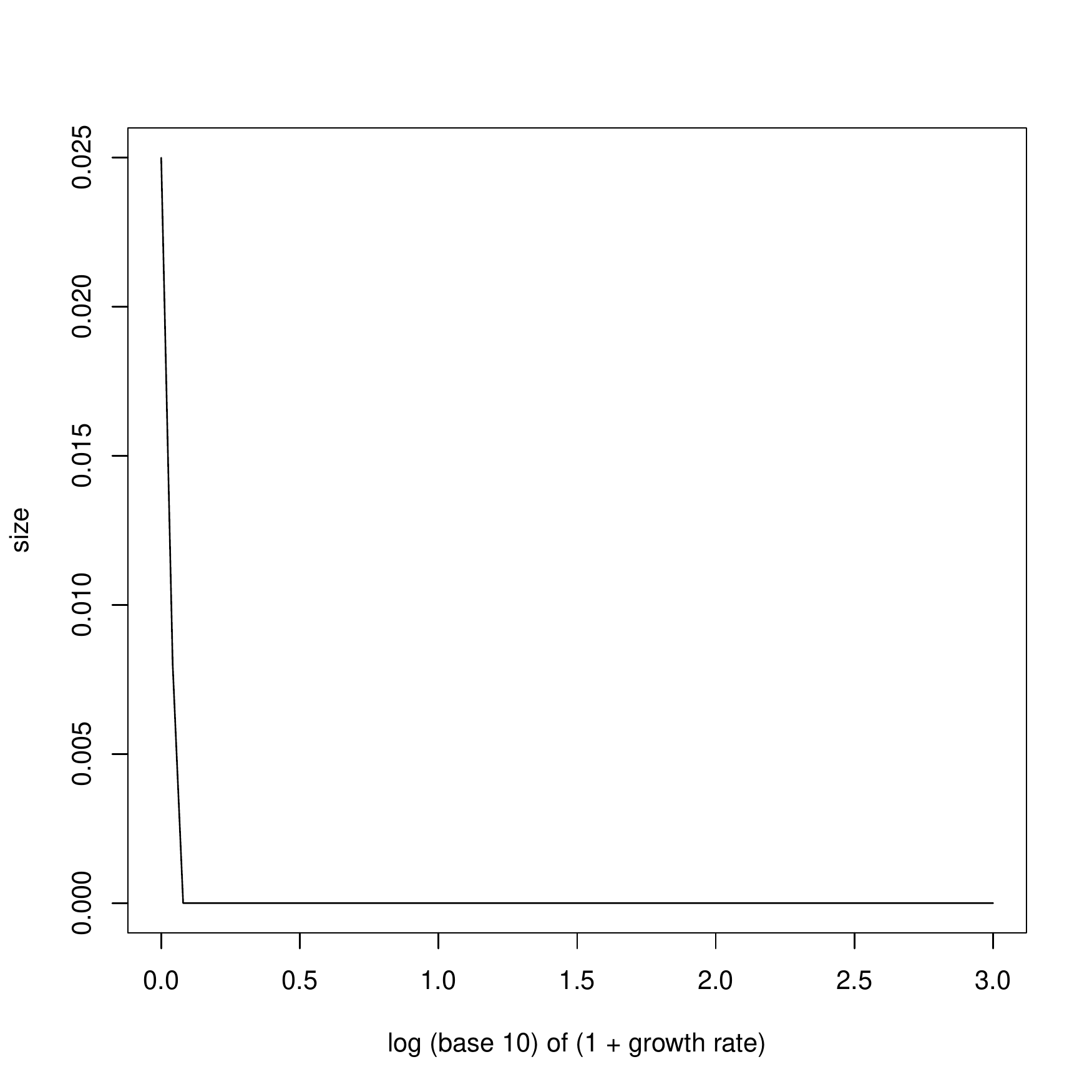}
\includegraphics[width = 0.49 \linewidth]{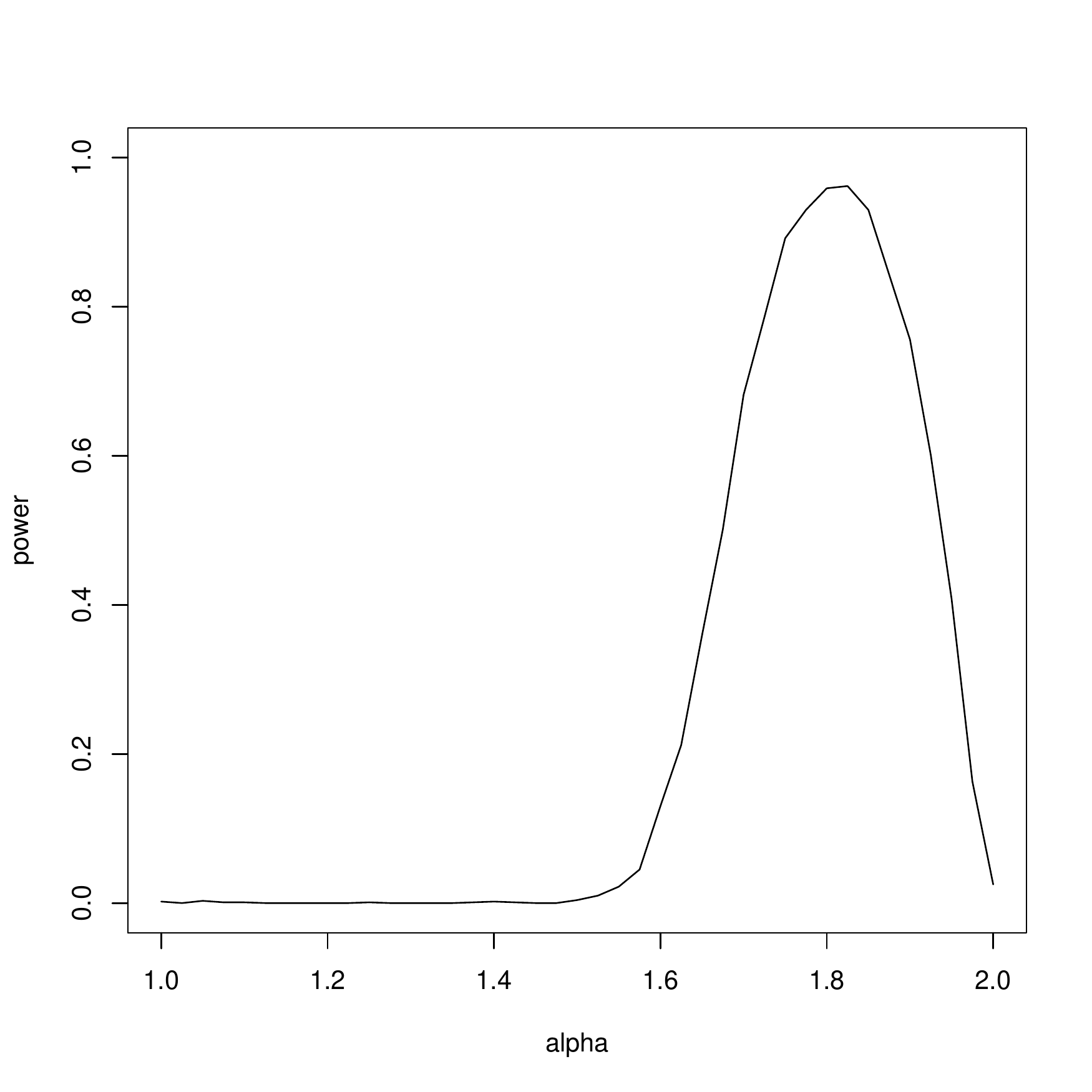}
\caption{(Top Row) Empirical size (Left) and power (Right) of a test of $\Theta_0$ vs $\Theta_1$ conducted using calibration data from panmictic models, but applied to pseudo-observed data from four deme models.
(Bottom Row) Empirical size (Left) and power (Right) of a test of $\Theta_0$ vs $\Theta_1$ conducted using calibration data from four deme models, but applied to pseudo-observed data from two deme models. 
The simulation parameters in both cases are as in Figure \ref{structure}.}
\label{test_migration}
\end{figure}

Figure \ref{structure} shows sampling distributions corresponding to a four deme model with symmetric migration between all pairs of demes, as well as a two deme model with asymmetric migration.
The contours differ markedly from the panmictic results in Figures \ref{selection} and \ref{recombination}, and also from each other.
Figure \ref{test_migration} demonstrates that misspecifying spatial structure results in very poor performance of the hypothesis test, with both the size and power curves showing complex behaviour that depends on the patterns of overlap between the distribution of the misspecified calibration data, and the pseudo-observed data.

\section{Distinguishing high fecundity from selective sweeps}\label{sweeps}

This section focuses on distinguishing multiple mergers due to selective sweeps from multiple mergers due to high fecundity.
The high fecundity model is the $\Xi$-coalescent introduced in Section \ref{alt_models} with a single deme ($D = 1$), no recombination $\rho = 0$, no selection $\sigma_i = 0$, and a constant population size $\lambda_1( t ) = 1$.
For selective sweeps, we assume a population of constant size evolving in non-overlapping generations, in which mutations providing a selective advantage $x \in ( 0, 1 )$ occur at points of a Poisson process with rate $x^{ -2 } \operatorname{Beta}( 2 - \alpha, \alpha )( dx )$, and sweep to fixation instantaneously on the coalescent time scale.

We also assume that recombination within chromosomes results in incomplete sweeps, so that when viewed backwards in time each individual has a random chance to participate in the merger resulting from each sweep.
Recombination is specified implicitly by setting the probability of a lineage participating in a sweep arising from a mutation with advantage $x \in ( 0, 1 )$ to $x$. 
Genetic material that is unlinked to the beneficial mutation escapes the selective sweep, and thus multiple mergers affect one chromosome at a time.
Neutral mutations continue to accrue along ancestral branches according to the infinite sites model with mutation rate $\theta > 0$.
When the population is diploid and biparental, these dynamics result in an ancestral process in which the marginal coalescent at each locus is the $\Xi$-coalescent with merger rates given by \eqref{xi_rate}.
\vskip 11pt
\begin{rmk}
The model described above has not been derived as a scaling limit of a finite population model of evolution.
Instead, it has been chosen to make the task of distinguishing between selective sweeps and high fecundity as difficult as possible.
For the same reason, we also scale the mutation rate as $\theta \propto \lim_{ N \rightarrow \infty } N^{ \alpha - 1 } \mu_N$ as in the model of \cite{S2003}.
For biological motivation, note that this model closely resembles the $\Lambda$-coalescent of \cite{DS05}, which was derived as a scaling limit of finite population models undergoing selective sweeps and recombination in much the same way as above.
However, their convergence result can only be used to obtain $\Lambda$-coalescents in which $\Lambda$ has an atom at 0, and hence it cannot be immediately used to obtain our model \citep[Example 2.5]{DS05}.
A model akin to ours could be obtained as a similar scaling limit by letting selective sweeps occur more frequently than the time scale of pairwise coalescence, thus causing the atom at 0 to vanish in the large population limit \citep{G00, DS05}.
\end{rmk}

We fix our null hypothesis as the class of selective sweep models described above with the parameter $\alpha \in ( 1, 2 )$ discretised as in \eqref{theta_1}.
The alternative hypothesis is the high fecundity $\Xi$-coalescent described at the beginning of this section.
The only difference between the two model classes is whether coalescence times at unlinked chromosomes are independent (under the selective sweep model), or positively correlated (under the high fecundity model).
The marginal coalescents at each chromosome coincide.
\begin{figure}[!ht]
\centering
\includegraphics[width = 0.49 \linewidth]{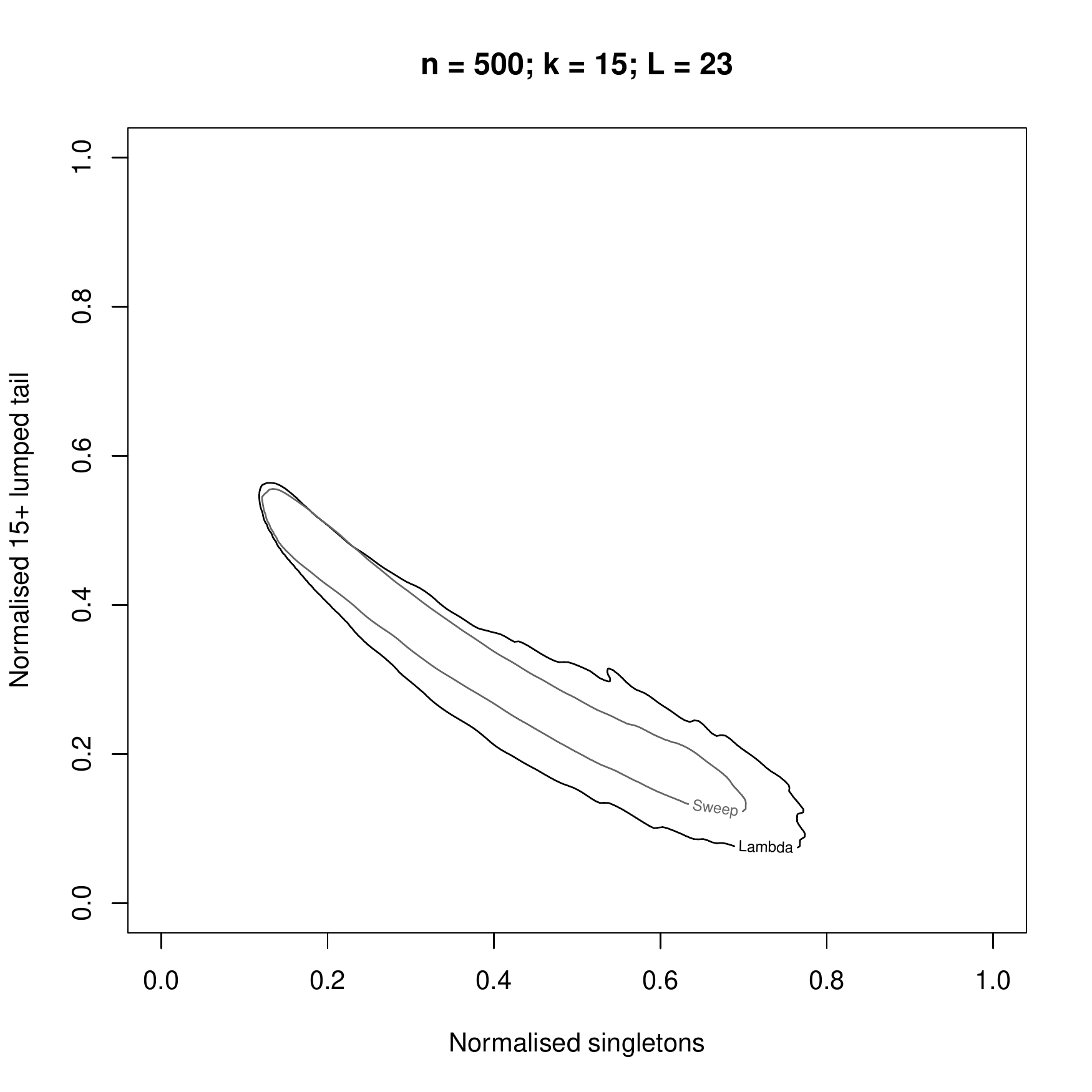}
\includegraphics[width = 0.49 \linewidth]{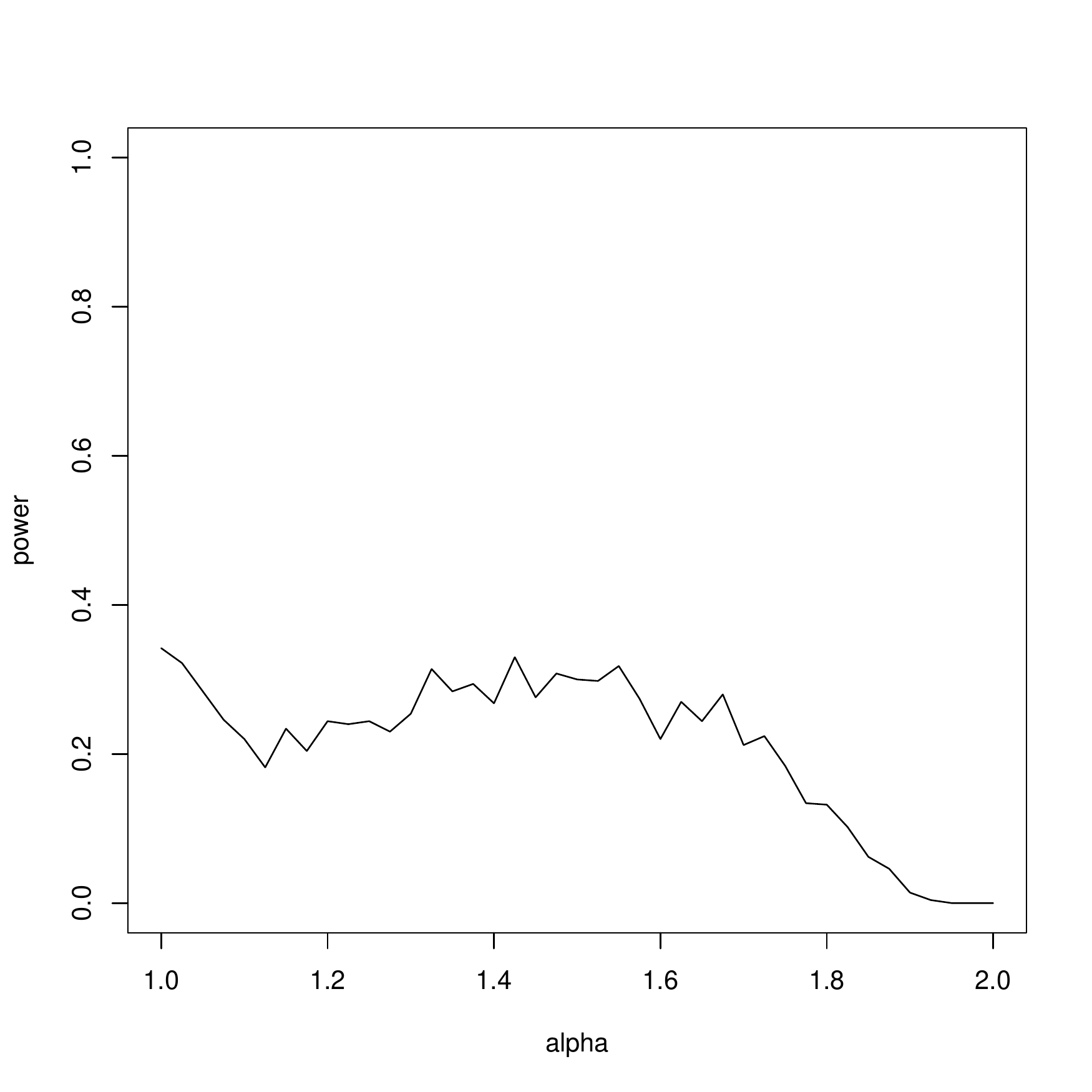} \\
\includegraphics[width = 0.49 \linewidth]{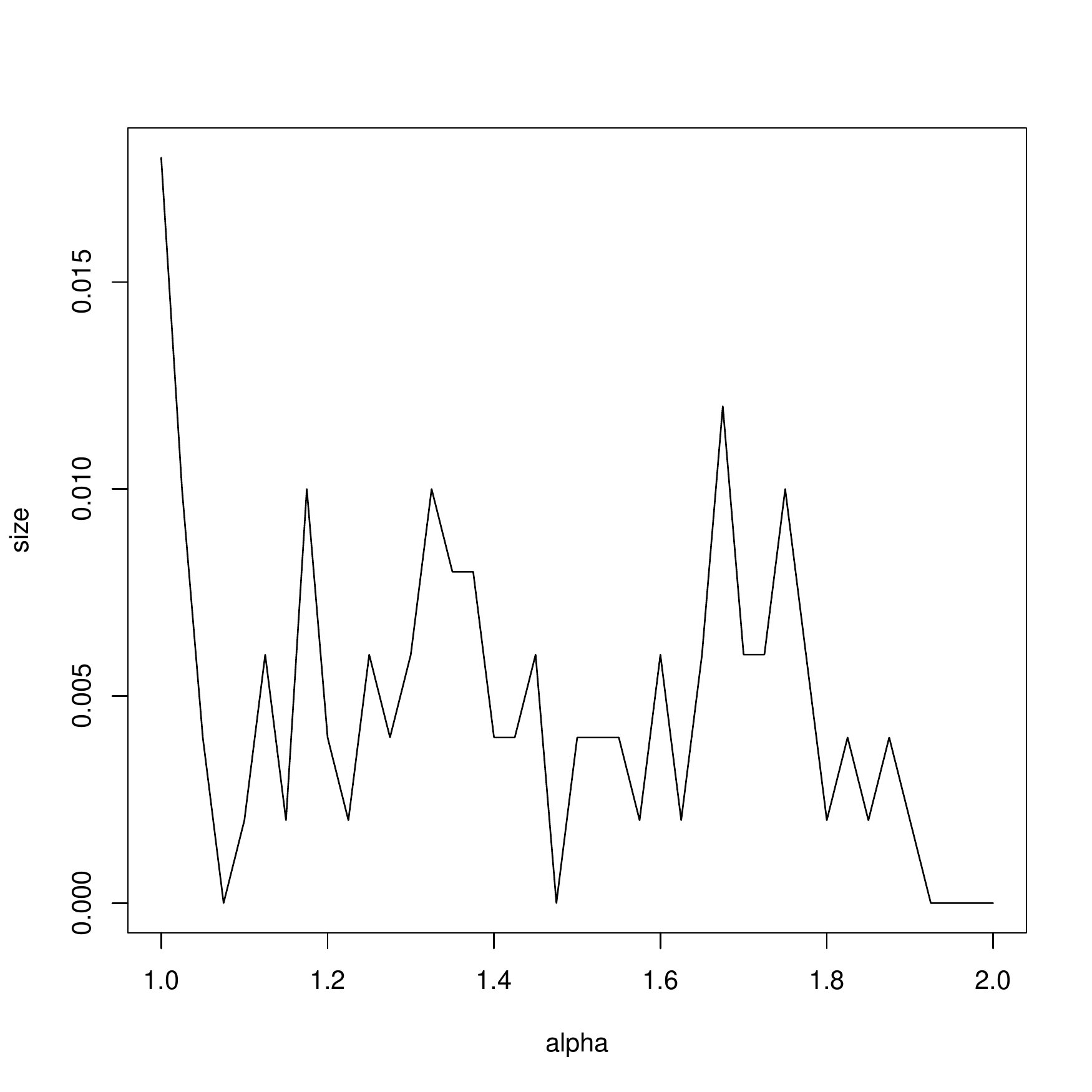}
\caption{(Top Left) $99^{\text{th}}$ percentile of a kernel density estimator fitted to 1000 realisations of the singleton-tail statistic under each model in $\Theta_0$ and $\Theta_1$. 
(Top Right) Empirical power and (Bottom) empirical size of the likelihood ratio test \eqref{test}.}
\label{ds}
\end{figure}

Figure \ref{ds} demonstrates that the singleton-tail statistic exhibits higher sampling variance under the alternative hypothesis than under the null due to positive correlation of coalescence times between unlinked chromosomes.
While the sampling distributions under both hypotheses centre on the same mean, increased variance means that the null hypothesis can be correctly rejected with moderate power of around $30\%$ for the majority of the parameter range.
Reversing the roles of the hypotheses caused the power to vanish, as the bulk of the sampling distribution under the alternative (selective sweep) hypothesis is fully contained in that of the null (high fecundity) hypothesis (results not shown).

\section{Discussion}\label{discussion}

We have derived a coalescent model of population growth and high fecundity involving multiple chromosomes under the joint effect of weak selection, recombination, and spatial structure.
We studied the effect of these three confounders on the ability of the singleton-tail statistic \cite{K18} to distinguish between population growth models and $\Xi$-coalescent models of high fecundity.

Crossover recombination and weak natural selection had no visible effect on the sampling distribution of the singleton-tail statistic.
Therefore, the statistic retained its ability to distinguish high fecundity from population growth with high power based on multi-locus data under these confounders.
Moreover, model selection can be based on calibration data simulated from a neutral model without recombination.
This reduces the number of nuisance parameters, and yields significant efficiency gains because selection and recombination are very expensive to simulate.
The computational speedup compensates for the relatively large sample size of 500, which appears to be necessary for a high-powered test, as samples of size 100 were shown in \cite{K18} to have noticeably lower power even without any of the confounders considered in this article.

Population structure had a significant effect on the sampling distribution of the singleton-tail statistic.
Misspecifying population structure when simulating calibration data rendered the hypothesis test inaccurate, with erratic  false positives and false negative probabilities.
It is well known that spatial structure results in an excess of intermediate and high frequency polymorphisms under the Kingman coalescent \citep{WA01, DD07}.
Our results confirm that similar phenomena also hold for $\Xi$-coalescents (Figure \ref{structure} shows a clear excess of high frequency polymorphisms and deficit of singletons), and that the exact amount of excess is sensitive to the details of the population structure.
This finding motivates research into methods which can infer population structure without assuming a particular coalescent or growth scenario.

Finally, we investigated the ability of the singleton-tail statistic to distinguish multiple mergers due to selective sweeps from multiple mergers due to high fecundity.
This is a challenging problem because the marginal models at single chromosomes coincide under the two hypotheses.
However, ancestral trees at unlinked chromosomes are independent under selective sweeps, which affect the genome locally, and positively correlated under high fecundity which affects all loci simultaneously (c.f.~\citep[Remark 1]{K18} for a formal justification).
Positive correlation increases the sampling variance of the multi-locus singleton-tail statistic, which enabled us to distinguish a high fecundity alternative hypothesis from a selective sweep null hypothesis with moderate power.
Reversing the roles of the hypotheses caused the power of the test to vanish, and thus model selection can only be successfully performed in one direction based on this method.

\section*{Acknowledgements}
The authors are grateful to Jochen Blath and Bjarki Eldon for discussions concerning generalised Beta-$\Xi$-coalescent models, and to Fabian Freund for pointing out an error in an earlier version of the manuscript. 
JK was supported by Deutsche Forschungsgemeinschaft (DFG) grant BL 1105/3-2 as part of SPP Priority Programme 1590, and by EPSRC grant EP/R044732/1. 
MWB was supported by DFG grant BL 1105/5-1 as part of SPP Priority Programme 1590. 
Both authors were also supported by DFG SPP Priority Programme 1819 start-up module grant ``Population genomics of highly fecund codfish".

\bibliographystyle{plainnat}
\bibliography{old_power}

\begin{thebibliography}{45}
\providecommand{\natexlab}[1]{#1}
\providecommand{\url}[1]{\texttt{#1}}
\expandafter\ifx\csname urlstyle\endcsname\relax
  \providecommand{\doi}[1]{doi: #1}\else
  \providecommand{\doi}{doi: \begingroup \urlstyle{rm}\Url}\fi

\bibitem[\'Arnason(2004)]{A2004}
E~\'Arnason.
\newblock Mitochondrial cytochrome \emph{b} variation in the high-fecundity
  {A}tlantic cod: trans-{A}tlantic clines and shallow gene genealogy.
\newblock \emph{Genetics}, 166:\penalty0 1871--1885, 2004.

\bibitem[Baake et~al.(2016)Baake, Lenz, and Wakolbinger]{BLW16}
E~Baake, U~Lenz, and A~Wakolbinger.
\newblock The common ancestor type distribution of a {$\Lambda$-Wright-Fisher}
  process with selection and mutation.
\newblock \emph{Electron Commun Probab}, 21\penalty0 (59):\penalty0 1--16,
  2016.

\bibitem[Beckenbach(1994)]{B1994}
A~T Beckenbach.
\newblock Mitochondrial haplotype frequencies in oysters: neutral alternatives
  to selection models.
\newblock In B~Golding, editor, \emph{Non-Neutral Evolution}, pages 188--198.
  Chapman \& Hall, New York, 1994.

\bibitem[Birkner et~al.(2011)Birkner, Blath, and Steinr\"ucken]{BBS2011}
M~Birkner, J~Blath, and M~Steinr\"ucken.
\newblock Importance sampling for {Lambda}-coalescents in the infinitely many
  sites model.
\newblock \emph{Theor Pop Biol}, 79:\penalty0 155--173, 2011.

\bibitem[Birkner et~al.(2013)Birkner, Blath, and Eldon]{BBE13a}
M~Birkner, J~Blath, and B~Eldon.
\newblock An ancestral recombination graph for diploid populations with skewed
  offspring distribution.
\newblock \emph{Genetics}, 193:\penalty0 255--290, 2013.

\bibitem[De and Durrett(2007)]{DD07}
A~De and R~Durrett.
\newblock Stepping-stone spatial structure causes slow decay of linkage
  disequilibrium and shifts the site frequency spectrum.
\newblock \emph{Genetics}, 176:\penalty0 969--981, 2007.

\bibitem[Donnelly and Kurtz(1999{\natexlab{a}})]{DK1999}
P~Donnelly and T~G Kurtz.
\newblock Particle representations for measure-valued population models.
\newblock \emph{Ann Probab}, 27:\penalty0 166--205, 1999{\natexlab{a}}.

\bibitem[Donnelly and Kurtz(1999{\natexlab{b}})]{DK99}
P~Donnelly and T~G Kurtz.
\newblock Genealogical processes for {Fleming-Viot} models with selection and
  recombination.
\newblock \emph{Ann Appl Probab}, 9:\penalty0 1091--1148, 1999{\natexlab{b}}.

\bibitem[Duong and Hazelton(2003)]{Duong03}
T~Duong and M~L Hazelton.
\newblock Plug-in bandwidth matrices for bivariate kernel density estimation.
\newblock \emph{J. Nonparametr. Statist.}, 15:\penalty0 17--30, 2003.

\bibitem[Durrett and Schweinsberg(2005)]{DS05}
R~Durrett and J~Schweinsberg.
\newblock A coalescent model for the effect of advantageous mutations on the
  genealogy of a population.
\newblock \emph{Stoch Proc Appl}, 115:\penalty0 1628--1657, 2005.

\bibitem[Eldon(2009)]{Eldon09}
B~Eldon.
\newblock Structured coalescent processes from a modified {Moran} model with
  large offspring numbers.
\newblock \emph{Theor Pop Biol}, 76:\penalty0 92--104, 2009.

\bibitem[Eldon and Wakeley(2006)]{EW2006}
B~Eldon and J~Wakeley.
\newblock Coalescent processes when the distribution of offspring number among
  individuals is highly skewed.
\newblock \emph{Genetics}, 172:\penalty0 2621--2633, 2006.

\bibitem[Eldon et~al.(2015)Eldon, Birkner, Blath, and Freund]{EBBF15}
B~Eldon, M~Birkner, J~Blath, and F~Freund.
\newblock Can the site frequency spectrum distinguish exponential population
  growth from multiple-merger coalescents.
\newblock \emph{Genetics}, 199\penalty0 (3):\penalty0 841--856, 2015.

\bibitem[Fearnhead(2001)]{F01}
P~Fearnhead.
\newblock Perfect simulation from population genetic models with selection.
\newblock \emph{Theor Pop Biol}, 59:\penalty0 263--279, 2001.

\bibitem[Fearnhead(2003)]{F03}
P~Fearnhead.
\newblock Ancestral process for non-neutral models of complex diseases.
\newblock \emph{Theor Pop Biol}, 63:\penalty0 115--130, 2003.

\bibitem[Fu(1995)]{Fu1995}
Y~X Fu.
\newblock Statistical properties of segregating sites.
\newblock \emph{Theor Pop Biol}, 48:\penalty0 172--197, 1995.

\bibitem[Gillespie(2000)]{G00}
J~H Gillespie.
\newblock Genetic drift in an infinite population: the pseudohitchhiking model.
\newblock \emph{Genetics}, 155:\penalty0 909--919, 2000.

\bibitem[Griffiths and Marjoram(1997)]{GM97}
R~C Griffiths and P~Marjoram.
\newblock An ancestral recombination graph.
\newblock In P~Donnelly and S~Tavar\'e, editors, \emph{Progress in population
  genetics and human evolution}, pages 257--270. Springer Verlag, Berlin, 1997.

\bibitem[Hedgecock and Pudovkin(2011)]{HP2011}
D~Hedgecock and A~I Pudovkin.
\newblock Sweepstakes reproductive success in highly fecund marine fish and
  shellfish: a review and commentary.
\newblock \emph{Bull Mar Sci}, 87:\penalty0 971--1002, 2011.

\bibitem[Herbots(1997)]{H97}
H~M Herbots.
\newblock The structured coalescent.
\newblock In P~Donnelly and S~Tavar\'e, editors, \emph{Progress in population
  genetics}, pages 231--255. Springer, New York, 1997.

\bibitem[Hudson(1983{\natexlab{a}})]{H1983a}
R~R Hudson.
\newblock Properties of a neutral allele model with intragenic recombination.
\newblock \emph{Theor Pop Biol}, 23:\penalty0 183--201, 1983{\natexlab{a}}.

\bibitem[Hudson(1983{\natexlab{b}})]{H1983b}
R~R Hudson.
\newblock Testing the constant-rate neutral allele model with protein sequence
  data.
\newblock \emph{Evolution}, 37:\penalty0 203--217, 1983{\natexlab{b}}.

\bibitem[Kingman(1982{\natexlab{a}})]{K82}
J~F~C Kingman.
\newblock The coalescent.
\newblock \emph{Stoch Proc Appl}, 13:\penalty0 235--248, 1982{\natexlab{a}}.

\bibitem[Kingman(1982{\natexlab{b}})]{K82b}
J~F~C Kingman.
\newblock Exchangeability and the evolution of large populations.
\newblock In G~Koch and F~Spizzichino, editors, \emph{Exchangeability in
  probability and statistics}, pages 97--112. North-Holland, Amsterdam,
  1982{\natexlab{b}}.

\bibitem[Kingman(1982{\natexlab{c}})]{Ki82c}
J~F~C Kingman.
\newblock On the genealogy of large populations.
\newblock \emph{J Appl Probab}, 19A:\penalty0 27--43, 1982{\natexlab{c}}.

\bibitem[Koskela(2018)]{K18}
J~Koskela.
\newblock Multi-locus data distinguishes between population growth and multiple
  merger coalescents.
\newblock \emph{Stat Appl Genet Mol Biol}, 17\penalty0 (3):\penalty0 20170011,
  2018.

\bibitem[Krone and Neuhauser(1997)]{KN97}
S~M Krone and C~Neuhauser.
\newblock Ancestral processes with selection.
\newblock \emph{Theor Pop Biol}, 51:\penalty0 210--237, 1997.

\bibitem[Limic and Sturm(2006)]{VL06}
V~Limic and A~Sturm.
\newblock The spatial {$\Lambda$}-coalescent.
\newblock \emph{Electron. J. Probab.}, 11:\penalty0 363--393, 2006.

\bibitem[Matuszewski et~al.(2018)Matuszewski, Hildebrandt, Achaz, and
  Jensen]{MHAJ18}
S~Matuszewski, M~E Hildebrandt, G~Achaz, and J~D Jensen.
\newblock Coalescent processes with skewed offspring distributions and
  non-equilibrium demography.
\newblock \emph{Genetics}, 208\penalty0 (1):\penalty0 1323--1338, 2018.

\bibitem[M\"ohle(1998)]{M98}
M~M\"ohle.
\newblock A convergence theorem for {Markov} chains arising in population
  genetics and the coalescent with selfing.
\newblock \emph{Adv in Appl Probab}, 30:\penalty0 493--512, 1998.

\bibitem[M\"ohle(2002)]{M2002}
M~M\"ohle.
\newblock The coalescent in population models with time-inhomogeneous
  environment.
\newblock \emph{Stochastic Process Appl}, 97:\penalty0 199--227, 2002.

\bibitem[M\"ohle and Sagitov(2001)]{MS2001}
M~M\"ohle and S~Sagitov.
\newblock Classification of coalescent processes for haploid exchangeable
  coalescent processes.
\newblock \emph{Ann Probab}, 29:\penalty0 1547--1562, 2001.

\bibitem[M\"ohle and Sagitov(2003)]{MS2003}
M~M\"ohle and S~Sagitov.
\newblock Coalescent patterns in diploid exchangeable population models.
\newblock \emph{J Math Biol}, 47:\penalty0 337--352, 2003.

\bibitem[Neuhauser and Krone(1997)]{NK97}
C~Neuhauser and S~M Krone.
\newblock The genealogy of samples in models with selection.
\newblock \emph{Genetics}, 145:\penalty0 519--534, 1997.

\bibitem[Pitman(1999)]{P1999}
J~Pitman.
\newblock Coalescents with multiple collisions.
\newblock \emph{Ann Probab}, 27:\penalty0 1870--1902, 1999.

\bibitem[Sagitov(1999)]{S1999}
S~Sagitov.
\newblock The general coalescent with asynchronous mergers of ancestral lines.
\newblock \emph{J Appl Probab}, 36:\penalty0 1116--1125, 1999.

\bibitem[Sargsyan and Wakeley(2008)]{SW2008}
O~Sargsyan and J~Wakeley.
\newblock A coalescent process with simultaneous multiple mergers for
  approximating the gene genealogies of many marine organisms.
\newblock \emph{Theor Pop Biol}, 74:\penalty0 104--114, 2008.

\bibitem[Schweinsberg(2000)]{S2000}
J~Schweinsberg.
\newblock Coalescents with simultaneous multiple collisions.
\newblock \emph{Electron J Prob}, 5:\penalty0 1--50, 2000.

\bibitem[Schweinsberg(2003)]{S2003}
J~Schweinsberg.
\newblock Coalescent processes obtained from supercritical {Galton-Watson}
  processes.
\newblock \emph{Stoch Proc Appl}, 106:\penalty0 107--139, 2003.

\bibitem[Steinr\"ucken et~al.(2013)Steinr\"ucken, Birkner, and Blath]{SBB2013}
M~Steinr\"ucken, M~Birkner, and J~Blath.
\newblock Analysis of {DNA} sequence variation within marine species using
  beta-coalescents.
\newblock \emph{Theor Pop Biol}, 87:\penalty0 15--24, 2013.

\bibitem[Tajima(1983)]{T1983}
F~Tajima.
\newblock Evolutionary relationship of {DNA} sequences in finite populations.
\newblock \emph{Genetics}, 105:\penalty0 437--460, 1983.

\bibitem[Tellier and Lemaire(2014)]{TL14}
A~Tellier and C~Lemaire.
\newblock Coalescence 2.0: a multiple branching of recent theoretical
  developments and their applications.
\newblock \emph{Mol Ecol}, 23:\penalty0 2637--2652, 2014.

\bibitem[T{\o}rresen et~al.(2017)T{\o}rresen, Star, Jentoft, Reinar, Grove,
  Miller, Walenz, Knight, Ekholm, Peluso, Edvardsen, {Tooming-Klunderud},
  Skage, Lien, Jakobsen, and Nederbragt]{T17}
O~K T{\o}rresen, B~Star, S~Jentoft, W~B Reinar, H~Grove, J~R Miller, B~P
  Walenz, J~Knight, J~M Ekholm, P~Peluso, R~B Edvardsen, A~{Tooming-Klunderud},
  M~Skage, S~Lien, K~S Jakobsen, and A~J Nederbragt.
\newblock An improved genome assembly uncovers prolific tandem repeats in
  {Atlantic} cod.
\newblock \emph{BMC Genomics}, 18\penalty0 (1):\penalty0 95, 2017.

\bibitem[Wakeley and Alicar(2001)]{WA01}
J~Wakeley and N~Alicar.
\newblock Gene genealogies in a metapopulation.
\newblock \emph{Genetics}, 159:\penalty0 893--905, 2001.

\bibitem[Watterson(1975)]{W75}
G~A Watterson.
\newblock On the number of segregating sites in genetical models without
  recombination.
\newblock \emph{Theor Pop Biol}, 7:\penalty0 1539--1546, 1975.

\end{thebibliography}

\end{document}